\documentclass[a4paper,12pt,parskip=half,english,numbers=noenddot]{scrartcl}

\usepackage{babel}
\usepackage[T1]{fontenc}
\usepackage[utf8]{inputenc}   
\usepackage{lmodern}
\usepackage{textcomp,latexsym}
\usepackage{verbatim}
\usepackage{cancel}
\usepackage{geometry}
\DeclareOldFontCommand{\it}{\normalfont\itshape}{\mathit}

\usepackage{amsmath,amssymb,amsthm}
\usepackage{bm}
\usepackage{bbm}

\usepackage{paralist}
\usepackage{enumerate}
\usepackage{threeparttable}
\usepackage{longtable,booktabs,multirow}
\usepackage{tabularx}
\usepackage{nth}
\usepackage{alphalph}
\usepackage{diagbox}

\usepackage[footnotesize,labelfont={bf}]{caption}
\usepackage[normalem]{ulem}
\usepackage{upgreek}

\usepackage{graphicx}
\usepackage{xcolor}
\graphicspath{{./pictures/}}
\usepackage{subcaption}

\usepackage{tikz}
\usetikzlibrary{
  decorations.pathmorphing,
  matrix,
  arrows,
  arrows.meta,
  calc,
  shapes,
  shapes.geometric
}

\usepgfmodule{nonlineartransformations}
\usepgflibrary{curvilinear}

\usepackage{tikz-3dplot}

\tikzstyle{block}     = [draw,rectangle,thick,minimum height=2em,minimum width=2em]
\tikzstyle{sum}       = [draw,circle,inner sep=0mm,minimum size=2mm]
\tikzstyle{connector} = [->,thick]
\tikzstyle{line}      = [thick]
\tikzstyle{branch}    = [circle,inner sep=0pt,minimum size=1mm,fill=black,draw=black]
\tikzstyle{guide}     = []
\tikzstyle{snakeline} = [
  connector,
  decorate,
  decoration={
    pre length=0.2cm,
    post length=0.2cm,
    snake,
    amplitude=.4mm,
    segment length=2mm
  },
  thick, magenta, ->
]

\usepackage[hidelinks]{hyperref}
\usepackage[boxed]{algorithm2e}

\DeclareCaptionStyle{table_new}{font={footnotesize},justification=raggedright}

\setdefaultenum{1)}{\theenumi.1)}{}{}

\makeatletter
\renewcommand*{\env@matrix}[1][*\c@MaxMatrixCols c]{%
  \hskip -\arraycolsep
  \let\@ifnextchar\new@ifnextchar
  \array{#1}}
\makeatother

\newtheorem{remark}{Remark}

\newcommand{\op}[1]{\operatorname{#1}}

\usepackage{scalerel}

\renewcommand{\vec}{\boldsymbol}   

\newcommand{\Div}{\operatorname{Div}}

\newcommand{\SymM}[1][3]{\op{S}^{#1}}

\newcommand{\tp}{^{\scriptstyle \mathrm{T}}}

\renewcommand{\det}{\operatorname{det}}

\newcommand{\mi}[1]{\vec{#1}}

\begin{document}

\begin{center}
\large{\textbf{Monolithic solution and null-space condensation of internal variables in finite viscoelasticity.}}

{\large A.K.K. Sethuraman, C. Hesch\footnote{Corresponding author. E-mail address: christian.hesch@uni-siegen.de}}

{\small
Chair of Computational Mechanics, University of Siegen, Siegen, Germany}

\end{center}

\begin{center}
\small
Published in \emph{Acta Mechanica}.
\href{https://doi.org/10.1007/s00707-026-04891-3}
{doi:10.1007/s00707-026-04891-3}
\end{center}

\vspace*{-0.1cm}\textbf{Abstract}

Finite-strain viscoelasticity is commonly discussed from a constitutive perspective, whereas the nonlinear solution architecture induced by time and space discretization is studied less systematically. In this work, we consider a representative viscoelastic model with a strain-like internal variable and focus on the fully discrete coupled problem in the deformation and the internal state. Starting from the underlying energy–dissipation structure, we derive the discrete weak forms and obtain a monolithic Newton system with a naturally non-symmetric block tangent. The increment of the internal variable is eliminated consistently at the level of the linearized system by a Schur complement, and the same reduction is shown to admit a geometric interpretation in terms of a null-space basis of the tangent space of the internal constraint manifold. This yields a genuinely monolithic counterpart to classical nested Gauss-point condensation, in which local constitutive equations are solved separately before the global equilibrium step. Numerical results for two- and three-dimensional Cook’s membrane benchmarks show that the proposed strategy retains essentially the same outer Newton behavior as the classical approach while substantially reducing computational cost by avoiding repeated local Newton solves; it also remains convergent for load increments for which the nested scheme fails. Moreover, the numerical study suggests that suitably chosen approximation spaces for the internal variable can yield additional savings in computational effort without significant loss of accuracy. Although presented for finite viscoelasticity, the construction extends naturally to broader classes of thermodynamically consistent internal-variable models.

\section{Introduction}

Finite-strain viscoelasticity has given rise to a broad range of constitutive models over the past decades \cite{lubliner_model_1985,simo_fully_1987,le_tallec_three-dimensional_1993,reese_theory_1998,bonet_large_2001}. A large part of the literature is driven by thermodynamic admissibility, constitutive flexibility, and the ability to reproduce experimental data \cite{lion_constitutive_1996,bergstrom_constitutive_1998,miehe_superimposed_2000,miehe_micromacro_2005,linder_micromechanically_2011,kleuter_generalized_2007}. From this perspective, the main emphasis is typically placed on the choice of free energy, on the formulation of the internal evolution law, and on the calibration of material parameters. Compared with the broad constitutive literature, the nonlinear solution architecture has received less systematic attention; representative contributions include \cite{hartmann_computation_2002,condemartin_energy-consistent_2014,netz_monolithic_2015,gros_energymomentum_2010,schiebl_structurepreserving_2021,liu_continuum_2021}. 

From a numerical point of view, however, this solution architecture is of central importance. After temporal and spatial discretization, finite viscoelasticity with internal variables leads to a coupled nonlinear algebraic system in the deformation and the internal state. In many implementations, the evolution equations for the internal variables are treated as local constitutive updates and are solved by nested Newton iterations at the quadrature-point level \cite{hartmann_computation_2002}. While this classical viewpoint is natural from the material-modeling side, it tends to hide the structure of the fully discrete problem and may lead to considerable computational overhead due to repeated local solves. Moreover, the interaction between the global equilibrium iterations, the local constitutive iterations, and the approximation space chosen for the internal variables is rarely studied systematically.

In the present contribution, we deliberately take the opposite viewpoint. We use a finite viscoelastic model with a strain-like internal variable $\vec{C}_i$ as a representative test bed, but our primary interest is not the proposal of yet another constitutive law. Instead, we focus on the common algorithmic structure shared by a broader class of thermodynamically consistent internal-variable models \cite{hackl_generalized_1997,ortiz_variational_1999,de_angelis_internal_2000,nedjar_frameworks_2002,nedjar_frameworks_2002-1}. The essential ingredients are a free energy, a local evolution equation (or, more generally, a set of internal constraints), and the availability of a consistent tangent for the internal residual. Once these ingredients are available, the coupled problem can be formulated and solved monolithically at the fully discrete level.

Starting from the underlying energy-dissipation structure of the continuum model, we derive weak forms of the balance of linear momentum and of the internal evolution equation and discretize them in time by a midpoint rule. This leads to a single nonlinear system in the deformation and the internal variable field. Its Newton linearization has a natural block structure and, in general, a non-symmetric tangent matrix. Based on this structure, the increment of the internal variable can be condensed systematically by a Schur complement. At the same time, the same reduction admits a geometric interpretation through a null-space basis of the tangent space of the internal constraint manifold. Related monolithic reduction ideas have already been shown to be effective in other nonlinear mechanics settings, such as computational homogenization \cite{hesch_variational_2024}, overlapping domain decomposition \cite{khristenko_multidimensional_2021}, and higher-order patch-coupling procedures \cite{dittmann_weak_2019}. These results motivate the present transfer of the methodology to finite viscoelasticity.

This viewpoint also clarifies the distinction between the proposed monolithic strategy and the classical nested Gauss-point condensation. In the classical setting, the internal evolution equation is solved locally and nonlinearly before the global equilibrium problem is assembled. In the monolithic setting pursued here, the internal residual remains part of the global Newton process and is condensed only at the level of the linearized system. Algebraically, both approaches are closely related, but algorithmically the difference is substantial: the monolithic formulation avoids repeated local Newton solves and retains the effect of non-vanishing internal residuals during the global iterations.

The aim of this paper is therefore twofold. First, we formulate a monolithic solution framework for finite viscoelasticity with a strain-like internal variable and show how the internal equations can be condensed consistently at the level of the fully discrete Newton system, both algebraically and geometrically. Second, we assess the computational consequences of this formulation in representative benchmark problems. To this end, we compare the proposed monolithic strategy with a classical nested condensation scheme in two- and three-dimensional Cook’s membrane examples and study how different approximation spaces for the internal variable affect accuracy, computational cost, and nonlinear robustness.

Although the numerical examples are restricted to finite viscoelasticity, the construction itself is not tied to one particular viscosity law. Rather, it reflects a fully discrete solver structure that is relevant for a broader class of thermodynamically consistent internal-variable models, including generalized standard materials and viscoplastic regularizations of finite-strain plasticity. In this sense, the present work is intended as a contribution to nonlinear computational mechanics as much as to finite viscoelasticity itself.

Section 1.1 summarizes the notation used throughout the paper. Section 2 introduces the continuum model and its thermodynamic structure. Section 3 derives the weak forms and the midpoint time discretization. Section 4 presents the monolithic Newton system together with the Schur-complement and null-space condensation of the internal variable. Finally, Section 5 discusses the numerical examples.

\subsection{Definitions and notations.}\label{sec:def}

This section summarizes the notation used throughout the paper.  A contraction of two vectors will be understood as $[\vec{a}\cdot\vec{b}] = a_i\,b_i$,  where the Einstein summation convention on repeated indices is used.  For two second order tensors it holds $[\vec{A}\,\vec{B}]_{ij} = A_{ik}\,B_{kj}$ and the double contraction reads  $[\vec{A}:\vec{B}] = A_{ij}\, B_{ij}$.  Next, we define the gradient with respect to the reference \(\nabla_{\!X}(\bullet)\) of a vector field \(\vec{a}\) and of a second-order tensor field \(\vec{A}\) as
\begin{equation}
[\nabla_{\!X}(\vec{a})]_{iJ}
  = \frac{\partial a_i}{\partial X_J},
\qquad
[\nabla_{\!X}(\vec{A})]_{iJK}
  = \frac{\partial A_{iJ}}{\partial X_K}.
\end{equation}
For the divergence operator it follows
\begin{equation}
[\nabla_{\!X}\cdot\vec{A}]_{i}
  = \frac{\partial A_{iJ}}{\partial X_J},
\qquad
[\nabla_{\!X}\cdot\mathfrak{A}]_{iJ}
  = \frac{\partial \mathfrak{A}_{iJK}}{\partial X_K}.
\end{equation}
We work in a three-dimensional reference configuration
$\mathcal{B}_0 \subset \mathbb{R}^3$.
Second-order tensors are elements of $\mathbb{R}^{3\times 3}$.
We denote by
\begin{equation}
  \SymM
  :=
  \bigl\{
    \vec{A} \in \mathbb{R}^{3\times 3}
    \,\big|\,
    \vec{A} = \vec{A}^\mathsf{T}
  \bigr\}
  \qquad\text{and}\qquad
  \SymM[+]
  :=
  \bigl\{
    \vec{A} \in \SymM
    \,\big|\,
    \vec{A}\ \text{is positive definite}
  \bigr\}
\end{equation}
the space of symmetric and the open cone of symmetric positive-definite
second-order tensors, respectively. Whenever we write $\vec{A} : \mathcal{B}_0 \to \SymM[+]$, this means that
for each material point $\vec{X}\in\mathcal{B}_0$ the value $\vec{A}(\vec{X},t)$ is a symmetric positive-definite $3\times 3$ tensor.

\section{Preliminaries and problem description}

We start with a short summary of nonlinear continuum mechanics. In particular, we aim at finite viscoelasticity with a strain-like internal variable, considered as additional field defined on its reference configuration as bounded Lipschitz domain $\mathcal{B}_0\subset\mathbb{R}^3$ with boundary $\partial\mathcal{B}_0$ and outward unit normal $\vec{N}$. The actual configuration $\mathcal{B}\subset\mathbb{R}^3$ with boundary $\partial\mathcal{B}$ and outward unit normal $\vec{n}$ is related to the reference configuration by a deformation mapping $\vec{\varphi}:\mathcal{B}_0\rightarrow\mathbb{R}^3$, such that $\mathcal{B} = \vec{\varphi}(\mathcal{B}_0)$. Material points are labelled by $\vec{X}\in\mathcal{B}_0$ with corresponding actual position \(\vec{x} = \vec{\varphi}(\vec{X})\). 

We formulate the continuum model in three spatial dimensions, $\mathcal{B}_0 \subset \mathbb{R}^3$, but the structure carries over verbatim to two-dimensional settings ($d=2$), e.g.\ plane-strain problems, as will be shown in the examples.

\subsection{Kinematics and internal variable}

Within nonlinear continuum mechanics, the deformation gradient is a second-order tensor field, described by
\[
  \vec{F}: \mathcal{B}_0 \times [0,T] \to \mathbb{R}^{3\times 3},\; \vec{F} = \nabla_{\!X} (\vec{\varphi}), 
\]
and the right Cauchy--Green tensor is given by $\vec{C}: \mathcal{B}_0 \times [0,T] \to \SymM[+],\; \vec{C} = \vec{F}\tp \vec{F}$. Moreover, the internal variable field is given by
\[
  \vec{C}_i : \mathcal{B}_0 \times [0,T] \to \SymM[+].
\]
The use of internal state variables to represent irreversible material
processes is rooted in the thermodynamic framework of Coleman and Gurtin
\cite{coleman_thermodynamics_1967}; see also the continuum-mechanical
treatments in \cite{germain_continuum_1983,haupt_continuum_2002}.
Viscous deformation is often described by an internal deformation gradient
$\vec{F}_i$ via the formal multiplicative split
\[
  \vec{F} = \vec{F}_e \vec{F}_i,
\]where $\vec{F}_e$ denotes the elastic part and $\vec{F}_i$ the inelastic (viscous) part.
This motivates the internal strain-like variable
\[
  \vec{C}_i = \vec{F}_i\tp \vec{F}_i \in \SymM[+],
\]
and the elastic right Cauchy--Green tensor
\[
  \vec{F}_e\tp \vec{F}_e
  = \vec{F}_i^{-\tp} \,\vec{C}\,\vec{F}_i^{-1}.
\]
For the variationally consistent framework, it is convenient to introduce
the generally non-symmetric mixed tensor
\[
  \vec{C}_e=\vec{C}\vec{C}_i^{-1}.
\]
It is entirely represented in the reference configuration and is similar to
$\vec{F}_e^T\vec{F}_e$; hence both tensors possess the same principal invariants in the
isotropic setting.

\subsection{Free energy, stresses and driving forces}\label{sec:energy}

The specific Helmholtz free energy per reference volume is additively split into
an equilibrium part and a non-equilibrium part:
\begin{equation}
  \Psi(\vec{C}, \vec{C}_i)
  = \Psi_\mathrm{eq}(\vec{C})
  + \Psi_\mathrm{neq}(\vec{C}_e), 
  \label{eq:Psi_split_en}
\end{equation}
where we assume isotropy so that both parts depend only on the invariants of
their arguments (cf. \cite{reese_theory_1998}, \cite{gros_energymomentum_2010}).

The second Piola--Kirchhoff stress follows from the usual constitutive relation
\[
  \vec{S} = 2\,\frac{\partial \Psi}{\partial \vec{C}}
  = \vec{S}_\mathrm{eq} + \vec{S}_\mathrm{neq}.
\]
By the chain rule applied to \eqref{eq:Psi_split_en},
\begin{align}
  \vec{S}_\mathrm{eq}
  &= 2\,\frac{\partial \Psi_\mathrm{eq}}{\partial \vec{C}},\\[0.3em]
  \vec{S}_\mathrm{neq}
  &= 2\,\frac{\partial \Psi_\mathrm{neq}}{\partial \vec{C}_e}\,:\, \frac{\partial \vec{C}_e}{\partial \vec{C}}
   = 2\,\frac{\partial \Psi_\mathrm{neq}}{\partial \vec{C}_e}\, \vec{C}_i^{-1},
\end{align}
since $\partial \vec{C}_e/\partial \vec{C} = \vec{I}\boxtimes\vec{C}_i^{-1}$. It is convenient to introduce the mixed thermodynamic driving force associated with the
non-equilibrium part
\begin{equation}
  \vec{M} = \vec{S}_\mathrm{neq}\,\vec{C}
    = 2\,\frac{\partial \Psi_\mathrm{neq}}{\partial \vec{C}_e}\,\vec{C}_i^{-1}\,\vec{C},
    \label{eq:Mandel_def_en}
\end{equation}
which will play the role of thermodynamic driving force for the viscous
evolution. Using the chain rule, it is straightforward to rewrite the thermodynamic driving force as
\begin{equation}
  \vec{M} = -2\,\frac{\partial \Psi_\mathrm{neq}}{\partial \vec{C}_i}\,\vec{C}_i,
    \label{eq:Mandel_def_en_2}
\end{equation}
where the minus sign results from differentiating the inverse internal
variable in $\vec{C}_e=\vec{C}\vec{C}_i^{-1}$. 
Note that $\vec{M}$ is in general not symmetric and according to \cite{gros_energymomentum_2010}, it is an isotropic tensor function in a linear space $\mathbb{L}(T_X^*\mathcal{B}_0,T_X^*\mathcal{B}_0)$.

\subsection{Clausius--Duhem inequality and conjugate pair \texorpdfstring{$(\vec{M},\vec{L}_i)$}{(M,Li)}}
\label{sec:CD_continuous}

In Lagrangian form the Clausius--Duhem inequality reads
\begin{equation}
  \mathcal{D}_\mathrm{int}
  = \vec{S} : \tfrac12 \dot{\vec{C}} - \dot{\Psi}(\vec{C},\vec{C}_i) \;\ge 0.
  \label{eq:CD_general_en}
\end{equation}
Using \eqref{eq:Psi_split_en},
\[
  \dot{\Psi}
  = \frac{\partial \Psi_\mathrm{eq}}{\partial \vec{C}} : \dot{\vec{C}}
  + \frac{\partial \Psi_\mathrm{neq}}{\partial \vec{C}_e} : \dot{\vec{C}}_e,
\]
where the elastic strain measure evolves according to
\[
 \dot{\vec{C}}_e = \dot{\vec{C}}\,\vec{C}_i^{-1}
         + \vec{C}\,\frac{\mathrm d}{\mathrm dt}(\vec{C}_i^{-1}).
\]
Using the standard relation
\[
  \frac{\mathrm d}{\mathrm dt}(\vec{C}_i^{-1})
  = -\,\vec{C}_i^{-1}\dot{\vec{C}}_i\,\vec{C}_i^{-1},
\]
we obtain
\[
  \dot{\vec{C}}_e = \dot{\vec{C}}\,\vec{C}_i^{-1}
            - \vec{C}\,\vec{C}_i^{-1}\dot{\vec{C}}_i\,\vec{C}_i^{-1}.
\]
Substituting into \eqref{eq:CD_general_en} and exploiting
$\vec{S}_\mathrm{eq} = 2\,\partial\Psi_\mathrm{eq}/\partial \vec{C}$ yields
\begin{align}
  \mathcal{D}_\mathrm{int}
  &= \vec{S}_\mathrm{neq} : \tfrac12 \dot{\vec{C}}
     - \frac{\partial \Psi_\mathrm{neq}}{\partial \vec{C}_e} : 
       \bigl(\dot{\vec{C}}\,\vec{C}_i^{-1} - \vec{C}\,\vec{C}_i^{-1}\dot{\vec{C}}_i\,\vec{C}_i^{-1}\bigr)
     \nonumber\\[0.3em]
  &= \Bigl(
        \vec{S}_\mathrm{neq} : \tfrac12 \dot{\vec{C}}
        - \frac{\partial \Psi_\mathrm{neq}}{\partial \vec{C}_e} : \dot{\vec{C}}\,\vec{C}_i^{-1}
     \Bigr)
     + \frac{\partial \Psi_\mathrm{neq}}{\partial \vec{C}_e} :
       \vec{C}\,\vec{C}_i^{-1}\dot{\vec{C}}_i\,\vec{C}_i^{-1}.
\end{align}
The first bracket vanishes by the constitutive definition of
$\vec{S}_\mathrm{neq} =2\,\partial_{\vec{C}}\Psi_\mathrm{neq}$ together with the chain rule for
$\Psi_\mathrm{neq}(\vec{C}_e)$; see \cite{reese_theory_1998}. The remaining purely viscous contribution can be written as
\begin{equation}
  \mathcal{D}_\mathrm{int}
  = \vec{M} : \vec{L}_i^T,
  \label{eq:Dint_M_Li_en}
\end{equation}
where $\vec{L}_i = \frac12\vec{C}_i^{-1}\dot{\vec{C}}_i$ is a rate-of-deformation tensor again in the linear space $\mathbb{L}(T_X\mathcal{B}_0,T_X\mathcal{B}_0)$ associated with the
internal process, i.e.\ a suitable material representation of the internal velocity gradient. In other words, the dissipation density is the power-conjugate
product of driving force $\vec{M}$ and flux $\vec{L}_i$. This can be, regarding eq. \eqref{eq:Mandel_def_en_2}, rewritten as
\[
\mathcal{D}_\mathrm{int} = -\frac{\partial \Psi_\mathrm{neq}}{\partial \vec{C}_i}:\dot{\vec{C}}_i.
\]
Equation \eqref{eq:Dint_M_Li_en} is the key outcome of thermodynamics: the
internal variable enters the Clausius--Duhem inequality only through the
pair $(\vec{M},\vec{L}_i)$, and any constitutive choice must ensure
$\mathcal{D}_\mathrm{int} \ge 0$.
This representation in terms of thermodynamically conjugate forces and fluxes
is closely related to the internal-variable formulation of continuum
thermodynamics developed, among others, by Coleman and Gurtin
\cite{coleman_thermodynamics_1967} and Germain, Nguyen, and Suquet
\cite{germain_continuum_1983}.
\begin{remark}
Although $\vec{C}_i$ and $\dot{\vec{C}}_i$ are symmetric,
the associated viscous deformation rate
\(
  \vec{L}_i := \tfrac12\,\vec{C}_i^{-1}\dot{\vec{C}}_i
\)
is in general not symmetric. Hence $\vec{L}_i$ (and $\vec{M}$)
are to be interpreted as mixed tensors. If symmetry is required for the thermodynamic driving force, one could use
\[
\vec{\Sigma}:= -\frac{\partial \Psi_{\mathrm{neq}}}{\partial \vec{C}_i}
= \frac12\,\vec{M}\,\vec{C}_i^{-1},\quad\vec{\Sigma}
\in \SymM,
\]
instead of the thermodynamic driving force from the non-equilibrium part.
\end{remark}

\subsection{Variational structure and evolution law}

We now use \eqref{eq:Dint_M_Li_en}  to define a constitutive relation between the thermodynamic
driving force $\vec{M}$ and the viscous rate $\vec{L}_i$. The resulting structure belongs
to the class of generalized standard materials introduced by Halphen and
Nguyen \cite{halphen_sur_1975}; see also
\cite{germain_continuum_1983,hackl_generalized_1997,
lemaitre_mechanics_1990}. In this class, the reversible response is generated
by a free-energy potential, whereas the irreversible evolution is governed by
a convex dissipation potential or its dual.

\subsubsection{Linear viscous law and positivity}

A natural and widely used choice is a linear relation of the form
\begin{equation}
  \vec{M} = \mathbb{V} : \vec{L}_i^T,
  \label{eq:linear_visc_law_en}
\end{equation}
where $\mathbb{V}$ is a fourth-order viscosity tensor. For isotropic
materials we use the usual volumetric--deviatoric split
\begin{equation}
  \mathbb{V}
  = V_\mathrm{vol}\,\mathbb{P}_\mathrm{vol}
    + 2V_\mathrm{dev}\,\mathbb{P}_\mathrm{dev},
  \qquad V_\mathrm{vol}> 0,\; V_\mathrm{dev} > 0,
\end{equation}
where $\mathbb{P}_\mathrm{vol}:\mi A = \op{vol}(\mi A)$, $\op{vol}(\mi A) = \frac1d\op{tr}(\mi{A})\mi I$ and $\mathbb{P}_\mathrm{dev}:\mi A = \op{dev}(\mi A)$, $\op{dev}(\mi A) = \mi A - \op{vol}(\mi A)$.
Inserting \eqref{eq:linear_visc_law_en} into \eqref{eq:Dint_M_Li_en} gives
\begin{equation}
  \mathcal{D}_\mathrm{int}
  = \vec{M} : \vec{L}_i^T = \vec{L}_i^T : \mathbb{V} : \vec{L}_i^T
  \;\ge 0.
\end{equation}
Since $V_{\mathrm{vol}}>0$ and $V_{\mathrm{dev}}>0$, the viscosity operator
$\mathbb{V}$ is positive definite on the considered tensor space.
Consequently,
\[
  \mathcal D_{\mathrm{int}}\geq 0
\]
with equality only for a vanishing viscous rate.
In this form, the evolution law is simply the constitutive equation
\eqref{eq:linear_visc_law_en}. It prescribes the stress-like variable
$\vec{M}$ in terms of the rate $\vec{L}_i$.

\subsubsection{Maximum dissipation principle}

The following construction is a local constitutive variational principle in
the sense of generalized standard materials
\cite{halphen_sur_1975,hackl_generalized_1997}. It should be
distinguished from a global Ritz principle for the coupled boundary-value
problem. Closely related incremental variational formulations of constitutive
updates and standard dissipative solids can be found in
\cite{ortiz_variational_1999,miehe_homogenization_2002,
miehe_multi-field_2011}. We start from the quadratic
dissipation potential in the rate $\vec{L}_i$,
\begin{equation}
  \Phi(\vec{L}_i) = \tfrac12\,\vec{L}_i : \mathbb{V} : \vec{L}_i.
\end{equation}
With this at hand, a variational principle can be formulated by introducing 
\begin{equation}
 \mathcal{J}(\vec{L}_i)
  := \vec{M}^T : \vec{L}_i - \Phi(\vec{L}_i)
  = \vec{M}^T : \vec{L}_i - \tfrac12\,\vec{L}_i : \mathbb{V} : \vec{L}_i.
\end{equation}
As $\Phi(\vec{L}_i)$ is convex, the functional $\mathcal{J}$ is concave in
$\vec{L}_i$, and the physically realized rate is therefore defined by
\[
  \vec{L}_i = \arg\sup_{\vec{L}_i}\mathcal{J}(\vec{L}_i).
\]
Stationarity yields
\begin{equation}\label{eq:evol_orig}
  \delta \mathcal{J}
  = \bigl(\vec{M}^T - \mathbb{V} : \vec{L}_i\bigr) : \delta\vec{L}_i
  = 0
  \quad\forall\,\delta\vec{L}_i,
\end{equation}
evaluated for a fixed $\vec{M}$. This local constitutive potential structure should not be confused with a
global Ritz formulation of the coupled initial-boundary-value problem.
Nevertheless, the constitutive equation can be embedded into the global weak
form by testing it with an admissible tensor-valued test function.

In the following we assume that the viscosity tensor \(\mathbb{V}\) is invertible. With this, the associated complementary (Legendre-dual) dissipation potential in the driving
force $\vec{M}$ is
\begin{align}
&\Phi^\ast(\vec{M})
:= \sup_{\vec{L}_i}
\bigl( \vec{M}^T : \vec{L}_i - \Phi(\vec{L}_i) \bigr)\\
&= \Bigl[
    \vec{M}^T : \vec{L}_i
    - \tfrac12\,\vec{L}_i : \mathbb{V} : \vec{L}_i
  \Bigr]_{\vec{L}_i = \mathbb{V}^{-1}:\vec{M}^T}
= \tfrac12\,\vec{M}^T : \mathbb{V}^{-1} : \vec{M}^T.
\end{align}
where $\mathbb{V}^{-1}$ denotes the inverse of the fourth-order tensor $\mathbb{V}$. 
Using $\Phi^\ast$, the principle of maximum dissipation can be formulated as follows:
for a given viscous rate $\vec{L}_i$, the physically realized thermodynamic driving
force $\vec{M}$ maximizes
\begin{equation}
  \mathcal{J}^\ast(\vec{M};\vec{L}_i)
  := \vec{M}^T : \vec{L}_i - \Phi^\ast(\vec{M})
  = \vec{M}^T : \vec{L}_i - \tfrac12\,\vec{M}^T : \mathbb{V}^{-1} : \vec{M}^T
\end{equation}
with respect to $\vec{M}$ for fixed $\vec{L}_i$. Hence, stationarity with regard to $\vec{M}$ gives
\begin{equation}
  \delta \mathcal{J}^\ast
  = \bigl(\vec{L}_i - \mathbb{V}^{-1} : \vec{M}^T\bigr)^T : \delta\vec{M}
  = 0
  \quad\forall\,\delta\vec{M},
\end{equation}
and thus
\begin{equation}
  \vec{L}_i = \bigl(\mathbb{V}^{-1} : \vec{M}\bigr)^T,
 \end{equation}
where the last identity follows from the transposition invariance of the
isotropic viscosity operator.
Since $\Phi^\ast$ is strictly convex for a positive-definite viscosity
tensor $\mathbb{V}$, the functional $\mathcal J^\ast(\vec{M};\vec{L}_i )$ is strictly
concave in $\vec{M}$. Consequently, its stationary point is the unique maximum. In this dual formulation, the evolution
law appears as the Euler equation of a local variational problem in $\vec{M}$:
among all admissible driving forces $\vec{M}$ it selects the one that maximizes the
dissipation functional $\mathcal{J}^\ast(\vec{M};\vec{L}_i)$ for the given rate $\vec{L}_i$.

For the strictly convex quadratic potential considered here, the primal and
dual statements are equivalent. For more general inelastic evolution laws,
the relation between evolution equations generated by dissipation potentials
and formulations based on the principle of maximum dissipation requires
additional assumptions; see Hackl and Fischer
\cite{hackl_relation_2008}.

As we have a linear relation between the driving force $\vec{M}$ and the
rate $\dot{\vec{C}}_i$, we can solve for $\dot{\vec{C}}_i$ using
\begin{equation}
  \dot{\vec{C}}_i
  := 2\,\vec{C}_i\bigl(\mathbb{V}^{-1}:\vec{M}(\vec{C},\vec{C}_i)\bigr)^T.
  \label{eq:Ci_continuous}
\end{equation}
and we obtain
\begin{equation}\label{eq:evol}
  \bigl(\dot{\vec{C}}_i - 2\,\vec{C}_i(\mathbb{V}^{-1} : \vec{M})^T\bigr) : \delta\vec{\Sigma}
  = 0
  \quad\forall\,\delta\vec{\Sigma}\quad\text{with}\quad\delta\vec{\Sigma}\in \SymM.
\end{equation}
For the isotropic constitutive structure considered here, the evolution
equation preserves the symmetry of $\vec{C}_i$.
In the vicinity of thermodynamic equilibrium ($\vec{C}_e\approx\vec{I}$),
a standard linearization of \eqref{eq:Ci_continuous} yields the familiar
finite linear viscoelastic relation
\begin{equation}
  \dot{\vec{C}}_i \approx \frac{1}{\tau}\bigl(\vec{C} - \vec{C}_i\bigr),
\end{equation}
with an effective relaxation time $\tau$; see \cite{reese_theory_1998}
for details in the small-deviation regime.

\subsection{Global energy functional and dissipation balance}

We consider a body $\mathcal{B}_0\subset \mathbb{R}^3$ in the reference configuration with mass density
$\rho_0(\vec{X})$ and motion $\vec{\varphi}(\vec{X},t)$. The Lagrangian velocity is
$\vec{v} := \dot{\vec{\varphi}}$. The kinetic energy reads
\begin{equation}
  E_{\mathrm{kin}}(t)
  =
  \frac{1}{2}
  \int_{\mathcal{B}_0}
    \rho_0(\vec{X})\,
    \vec{v}(\vec{X},t)\cdot\vec{v}(\vec{X},t)\,
  \mathrm dV.
  \label{eq:T_global_en}
\end{equation}
The stored (Helmholtz) free energy of the viscoelastic material is obtained by spatial
integration of the local density $\Psi(\vec{C},\vec{C}_i)$,
\begin{equation}
  U_\mathrm{int}(t)
  :=
  \int_{\mathcal{B}_0}
    \Psi\bigl(\vec{C}(\vec{X},t),\vec{C}_i(\vec{X},t)\bigr)\,
  \mathrm dV,
  \label{eq:U_int_global_en}
\end{equation}
and, by definition, we obtain
\begin{equation}
  \dot{U}_\mathrm{int}(t)
  =
  \int_{\mathcal{B}_0}
    \dot{\Psi}(\vec{C},\vec{C}_i)\,
  \mathrm dV,
\end{equation}
where
\[
  \dot\Psi
  =
  \frac{\partial\Psi}{\partial\vec{C}}:\dot{\vec{C}}
  +
  \frac{\partial\Psi}{\partial\vec{C}_i}:\dot{\vec{C}_i}.
\]
The local Clausius--Duhem inequality \eqref{eq:CD_general_en} yields the global internal dissipation
\begin{equation}
  D_\mathrm{int}(t)
  :=
  \int_{\mathcal{B}_0} \mathcal{D}_\mathrm{int}\,\mathrm dV
  =
  \int_{\mathcal{B}_0}
    \vec{S} : \tfrac12 \dot{\vec{C}}\,
  \mathrm dV
  - \dot{U}_\mathrm{int}(t)
  \;\ge 0,
  \label{eq:D_int_global_en}
\end{equation}
or equivalently, using \eqref{eq:Dint_M_Li_en},
\begin{equation}
  D_\mathrm{int}(t)
  =
  \int_{\mathcal{B}_0}
    \vec{M}(\vec{X},t) : \vec{L}_i^T(\vec{X},t)\,
  \mathrm dV
  \;\ge 0.
  \label{eq:D_int_MLi_global_en}
\end{equation}
The internal mechanical power is given by
\begin{equation}
  P_\mathrm{int}(t)
  :=
  \int_{\mathcal{B}_0}
    \vec{S} : \tfrac12 \dot{\vec{C}}\,
  \mathrm dV,
  \label{eq:P_int_global_en}
\end{equation}
such that \eqref{eq:D_int_global_en} can be written in compact form
\begin{equation}
  \dot{U}_\mathrm{int}(t)
  =
  P_\mathrm{int}(t) - D_\mathrm{int}(t),
  \qquad D_\mathrm{int}(t)\ge 0.
  \label{eq:Udot_equals_Pint_minus_D_en}
\end{equation}
Let the boundary of $\mathcal{B}_0$ be decomposed into a Dirichlet part
$\partial\mathcal{B}_0^u$ and a Neumann part $\partial\mathcal{B}_0^t$, and let
$\vec{b}_0(\vec{X},t)$ and $\vec{t}_0(\vec{X},t)$ denote prescribed body forces and
tractions (per reference volume and per reference area, respectively). The external
mechanical power is
\begin{equation}
  P_\mathrm{ext}(t)
  :=
  \int_{\mathcal{B}_0}
    \rho_0\,\vec{b}_0\cdot\vec{v}\,\mathrm dV
  +
  \int_{\partial\mathcal{B}_0^t}
    \vec{t}_0\cdot\vec{v}\,\mathrm dA.
  \label{eq:P_ext_global_en}
\end{equation}
The balance of linear momentum, multiplied by $\vec{v}$ and integrated over
$\mathcal{B}_0$, yields the global mechanical power balance
\begin{equation}
  \dot{E}_{\mathrm{kin}}(t) + P_\mathrm{int}(t) = P_\mathrm{ext}(t).
  \label{eq:mech_power_balance_en}
\end{equation}

Combining \eqref{eq:Udot_equals_Pint_minus_D_en} and
\eqref{eq:mech_power_balance_en} leads to the global energy balance
\begin{equation}
  \frac{\mathrm d}{\mathrm dt}
  \Bigl( E_{\mathrm{kin}}(t) + U_\mathrm{int}(t) \Bigr)
  =
  P_\mathrm{ext}(t) - D_\mathrm{int}(t),
  \qquad D_\mathrm{int}(t)\ge 0.
  \label{eq:global_energy_balance_en}
\end{equation}
If all external loads are conservative, $P_\mathrm{ext}(t)$ can be written as the negative
time derivative of an external potential $U_\mathrm{ext}(t)$, and the total mechanical
energy
\begin{equation}
  \mathcal{V}(t)
  :=
  E_{\mathrm{kin}}(t) + U_\mathrm{int}(t) + U_\mathrm{ext}(t)
\end{equation}
satisfies
\begin{equation}\label{eq:power}
  \dot{\mathcal{V}}(t) = -\,D_\mathrm{int}(t) \;\le 0.
\end{equation}
Thus, the continuous formulation automatically complies with the
second law of thermodynamics: the total energy can only decay by the
amount of viscous dissipation. In the fully discrete setting the midpoint 
discretization is designed to mimic the underlying energy–dissipation structure
as closely as possible, see \cite{gros_energymomentum_2010} for the algorithmic enforcement of 
the energy balance in the discrete setting. 
At this point, the continuous thermodynamic structure of the model is
completely specified by the free energy $\Psi(\vec{C},\vec{C}_i)$, the
Clausius--Duhem inequality \eqref{eq:CD_general_en} and the evolution law
for $\vec{C}_i$.

\section{Integral weak forms and midpoint time discretization}
\label{sec:weak_forms_time}

Starting from the global energy and power balances derived above, we now
introduce weak forms of the momentum balance and of the evolution equation
on $\mathcal{B}_0$. These will be discretized in time by a midpoint rule in
order to obtain a fully discrete system that preserves the underlying
energy--dissipation structure as closely as possible.

\subsection{Weak form of the momentum balance}

We now elaborate the problem described in \eqref{eq:power} in detail. Starting from 
the balance of linear momentum in the reference configuration, multiplication by an 
admissible virtual displacement and integration over $\mathcal{B}_0$ yields the weak
form: Find $\vec{\varphi}(t)$ with
$\vec{\varphi}(t) = \bar{\vec{\varphi}}(t)$ on $\partial\mathcal{B}_0^\mathrm{D}$ and $\delta\vec{\varphi}(t) = \vec{0}$ on $\partial\mathcal{B}_0^\mathrm{D}$,
such that for all admissible test functions $\delta\vec{\varphi}$ with
\begin{equation}
  \int_{\mathcal{B}_0}
    \rho_0\,\delta\vec{\varphi}\cdot\ddot{\vec{\varphi}}\,\mathrm{d}V
  +
  \int_{\mathcal{B}_0}
    \vec{P}(\vec{C},\vec{C}_i) : \nabla_{\!X}(\delta\vec{\varphi})\,\mathrm{d}V
  =
  \int_{\mathcal{B}_0}
    \rho_0 \,\vec{b}_0 \cdot \delta\vec{\varphi}\,\mathrm{d}V
  +
  \int_{\partial\mathcal{B}_0^\mathrm{N}}
    \bar{\vec{t}}_0 \cdot \delta\vec{\varphi}\,\mathrm{d}A.
  \label{eq:weak_momentum}
\end{equation}
Here $\vec{P}$ is the first Piola–Kirchhoff stress induced by the second
Piola–Kirchhoff stress $\vec{S}(\vec{C},\vec{C}_i)$ from Section \ref{sec:energy}, i.e.\ $\vec{P} = \vec{F}\,\vec{S}$.
Using standard arguments in variational calculus, we end up with the local balance of linear momentum in the reference configuration
\[
  \Div_{\!X} \vec{P}(\vec{C},\vec{C}_i) + \rho_0 \,\vec{b}_0(\vec{X},t)
  = \rho_0\,\ddot{\vec{\varphi}}(\vec{X},t)
  \quad\text{in }\mathcal{B}_0,
\]
with prescribed tractions $\bar{\vec{t}}_0$ on $\partial\mathcal{B}_0^\mathrm{N}$ and prescribed motion $\bar{\vec{\varphi}}$ on $\partial\mathcal{B}_0^\mathrm{D}$. Note that this local form allows us to elaborate the explicit structure of the boundary conditions in detail, which is a non-trivial task for, e.g., higher-order materials.

For the weak form of the internal evolution equation, the local stationarity
condition \eqref{eq:evol} is integrated over the reference configuration with an arbitrary symmetric
tensor field $\delta\vec{\Sigma} : \mathcal{B}_0\to\SymM$,
\begin{equation}
  \int_{\mathcal{B}_0}
    \delta\vec{\Sigma} :
    \bigl(
      \dot{\vec{C}}_i
      - 2\,\vec{C}_i\mathbb{V}^{-1} : \vec{M}^T
    \bigr)\,\mathrm{d}V = 0.
  \label{eq:weak_evol_Ci}
\end{equation}
Alternatively, we may use $\delta\vec{C}_i : \mathcal{B}_0\to\SymM$ following the formulation presented in \eqref{eq:evol_orig}. 
In a classical quadrature-point implementation, the distinction between
testing with $\delta\vec{\Sigma}$ or an equivalent representation in terms of
$\delta\vec{C}_i$ is usually immaterial because the local evolution equation is
enforced pointwise. In the present field formulation, however, the chosen
trial and test spaces determine the resulting algebraic coupling and therefore
have to be specified explicitly.

\subsection{Temporal discretisation}
\label{sec:midpoint_evol_weak}

For the time-discrete problem we consider two consecutive time levels $t_n$ and $t_{n+1}$ with step size $h = t_{n+1} - t_n$. Let
\[
  \vec{v} := \dot{\vec{\varphi}},\qquad
  \vec{v}^n := \vec{v}(\cdot,t_n),\quad
  \vec{v}^{n+1} := \vec{v}(\cdot,t_{n+1}),
\]
and introduce midpoint quantities by
\[
  (\bullet)_{n+\frac12} := \tfrac12\bigl((\bullet)^{n+1} + (\bullet)^n\bigr),
  \qquad
  \vec{v}_{n+\frac12}
  := \frac{\vec{\varphi}^{n+1} - \vec{\varphi}^n}{h},\quad
  \vec{a}_{n+\frac12}
  := \frac{\vec{v}^{n+1} - \vec{v}^n}{h},
\]
such that
\[
\vec{\varphi}^{n+1} - \vec{\varphi}^n = \frac{h}{2}(\vec{v}^{n+1} + \vec{v}^n),
\]
providing a direct relationship between $\vec{\varphi}^{n+1}$ and $\vec{v}^{n+1}$ for given values at time $t = t_n$.
Next, we introduce midpoint kinematic and material quantities
\[
  \vec{C}_{n+\frac12}
  := \vec{F}_{n+\frac12}^\mathsf{T}\vec{F}_{n+\frac12},\qquad
  \vec{C}_{i,n+\frac12}
  := \tfrac12(\vec{C}_i^{\,n+1} + \vec{C}_i^n),
\]
with $\vec{F}_{n+\frac12} := \nabla_{\!X}(\vec{\varphi}_{n+\frac12})$ and
$\vec{C}_{e,n+\frac12}
  := \vec{C}_{n+\frac12}\,\vec{C}_{i,n+\frac12}^{-1}$.
The first Piola–Kirchhoff stress at the midpoint is then
\[
  \vec{P}_{n+\frac12}
  :=
  \vec{P}\bigl(\vec{C}_{n+\frac12},\vec{C}_{i,n+\frac12}\bigr),
\]
and the external data are evaluated at the midpoint, $\vec{b}_{0,n+\frac12} := \vec{b}_0(\cdot,t_{n+\frac12})$ and $\bar{\vec{t}}_{0,n+\frac12} := \bar{\vec{t}}_0(\cdot,t_{n+\frac12})$.
Choosing $\delta\vec{\varphi}$ piecewise constant on $[t_n,t_{n+1}]$, the weak form \eqref{eq:weak_momentum} leads to the time-discrete counterpart
\begin{equation}
  \int_{\mathcal{B}_0}
    \rho_0\,\delta\vec{\varphi}
    \cdot\vec{a}_{n+\frac12}\,\mathrm{d}V
  +
  \int_{\mathcal{B}_0}
    \vec{P}_{n+\frac12}
    : \nabla_{\!X}(\delta\vec{\varphi})\,\mathrm{d}V
  =
  \int_{\mathcal{B}_0}
    \rho_0 \,\vec{b}_{0,n+\frac12}
    \cdot \delta\vec{\varphi}\,\mathrm{d}V
  +
  \int_{\partial\mathcal{B}_0^\mathrm{N}}
    \bar{\vec{t}}_{0,n+\frac12}
    \cdot \delta\vec{\varphi}\,\mathrm{d}A.
  \label{eq:weak_momentum_mid}
\end{equation}
Equation \eqref{eq:weak_momentum_mid} is the midpoint-discrete weak
form of the momentum balance and will later give rise, after spatial
discretization, to the global equilibrium residual
$\vec{R}_q(\vec{q}^{\,n+1},\vec{C}_i^{\,n+1})$ in the monolithic
Newton system.

The corresponding midpoint thermodynamic driving force is obtained from the free energy as
\begin{equation}
  \vec{M}_{n+\frac12}
  :=
  \vec{M}\bigl(\vec{C}_{n+\frac12},\vec{C}_{i,n+\frac12}\bigr)
  =
  2\,
    \frac{\partial\Psi_\mathrm{neq}}{\partial\vec{C}_e}
    \Bigl(\vec{C}_{e,n+\frac12}\Bigr)
    \, \vec{C}_{i,n+\frac12}^{-1}\,\vec{C}_{n+\frac12}.
\end{equation}
Moreover, we approximate the rate of the internal variable by the difference quotient,
\[
  \dot{\vec{C}}_i(\cdot,t_{n+\frac12})
  \approx
  \frac{\vec{C}_i^{n+1} - \vec{C}_i^n}{h},
\]
Restricting the weak evolution equation \eqref{eq:weak_evol_Ci} to the interval $[t_n,t_{n+1}]$ gives
\begin{equation}
  \int_{\mathcal{B}_0}
    \delta\vec{\Sigma} :
    \Bigl(
      \frac{\vec{C}_i^{n+1} - \vec{C}_i^n}{h}
      - 2\,\vec{C}_{i,n+\frac12}\,\mathbb{V}^{-1} : \vec{M}_{n+\frac12}^T
    \Bigr)\,\mathrm{d}V
  = 0
  \quad\forall\,\delta\vec{\Sigma},
  \label{eq:weak_evol_midpoint_M}
\end{equation}where $\delta\vec{\Sigma}$ is also piecewise constant on $[t_n,t_{n+1}]$.

\subsection{Spatial finite element discretization}
\label{sec:spatial_FE}

So far, the weak forms \eqref{eq:weak_momentum_mid} and
\eqref{eq:weak_evol_midpoint_M} are continuous in space. We now
introduce a standard finite element approximation on the reference
configuration $\mathcal{B}_0$. 
Let $\mathcal{B}_0$ be decomposed into finite elements
$\mathcal{B}_0^e$ with nodal points $a\in\mathcal{I}^e$, and let
$N_a(\vec{X})$ denote the usual $C^0$ Lagrange shape functions
associated with the displacement field. The discrete motion at time
$t_n$ is approximated by
\[
  \delta\vec{\varphi}_h(\vec{X})
  = \sum_{a\in\mathcal{I}} N_a(\vec{X})\,\delta\vec{q}_a,\qquad
  \vec{\varphi}_h^n(\vec{X})
  = \sum_{b\in\mathcal{I}} N_b(\vec{X})\,\vec{q}_b^n,
\]
where $\vec{q}_b^n$ are the nodal positions (or displacements) and
$\delta\vec{q}_a$ are the corresponding virtual nodal variations.
The midpoint kinematics follow from the same interpolation,
\[
  \vec{\varphi}_{h,n+\frac12}
  := \tfrac12(\vec{\varphi}_h^{n+1} + \vec{\varphi}_h^n),\qquad
  \vec{F}_{h,n+\frac12}
  := \nabla_{\!X}(\vec{\varphi}_{h,n+\frac12}),\qquad
  \vec{C}_{h}^{n+\frac12}
  := \vec{F}_{h,n+\frac12}^\mathsf{T}\vec{F}_{h,n+\frac12}.
\]
For the internal variable and its thermodynamic driving force we introduce,
in general, an approximation space with shape functions $H_\alpha(\vec{X})$:
\[
  \delta\vec{\Sigma}_h(\vec{X})
  = \sum_{\alpha\in\mathcal{J}} H_\alpha(\vec{X})\,\delta\vec{\Sigma}_\alpha,
\]
and
\[
  \vec{C}_{i,h}^{n+\frac12}(\vec{X})
  = \sum_{\beta\in\mathcal{J}} H_\beta(\vec{X})\,\vec{C}_{i,\beta}^{n+\frac12}.
\]
The same basis $H_\beta$ is used for $\vec{C}_{i,h}^{n+\frac12}$ and $\delta\vec{\Sigma}_h$ for
notational convenience; in particular, it may be chosen elementwise discontinuous as we do not require $C^0$ continuity. The full discrete elastic measure reads
\[
  \vec{C}_{e,h}^{n+\frac12}
  := \vec{C}_{h}^{n+\frac12}\,\vec{C}_{i,h}^{-1,n+\frac12},
 \]
and the discrete thermodynamic driving force $\vec{M}_{h}^{n+\frac12}  = \vec{M}\bigl(\vec{C}_{h}^{n+\frac12},\vec{C}_{i,h}^{n+\frac12}\bigr)$ follows from the free energy as presented in Section \ref{sec:energy}.

\color{black}
In the following, the superscript $n+1$ attached to the residuals denotes
the nonlinear algebraic step problem for the unknown endpoint values
$q^{n+1}$ and $C_i^{n+1}$, with the state at $t_n$ kept fixed. It should
not be read as a pointwise evaluation of all quantities at $t_{n+1}$.
Unless stated otherwise, the constitutive quantities entering the
residuals are evaluated at the midpoint state introduced above.

\color{black}

\paragraph{Full-discrete momentum equation.}

Inserting $\vec{\varphi}_h$, $\delta\vec{\varphi}_h$ and the midpoint
quantities into \eqref{eq:weak_momentum_mid} and exploiting the
arbitrariness of the nodal variations $\delta\vec{q}_a$ yields the
full-discrete balance of momentum 
\begin{equation}
  \sum_{a\in\mathcal{N}}
    \delta\vec{q}_a \cdot
    \vec{R}_{q,a}^{\,n+1}
  = 0
  \quad\forall\,\{\delta\vec{q}_a\},
\end{equation}
with nodal residuals
\begin{align}\label{eq:residual}
  \vec{R}_{q,a}^{\,n+1}
  &:=
  \int_{\mathcal{B}_0}
    \rho_0\,N_a\,\vec{a}_{n+\frac12}\,\mathrm{d}V
  +
  \int_{\mathcal{B}_0}
    \vec{P}_{h}^{n+\frac12}
    : \nabla_{\!X}(N_a)\,\mathrm{d}V
\nonumber\\[0.3em]
  &\quad
  -
  \int_{\mathcal{B}_0}
    \rho_0\,N_a\,\vec{b}_{0,n+\frac12}\,\mathrm{d}V
  -
  \int_{\partial\mathcal{B}_0^\mathrm{N}}
    N_a\,\bar{\vec{t}}_{0,n+\frac12}\,\mathrm{d}A,
\end{align}
where $\vec{P}_{h}^{n+\frac12} := \vec{F}_{h,n+\frac12}\vec{S}(\vec{C}_{h}^{n+\frac12},\vec{C}_{i,h}^{n+\frac12})$.

Collecting the nodal residuals in a global vector $\vec{R}_q^{\,n+1}$ and the nodal unknowns in $\vec{q}^{\,n+1}$ for every node $a\in\mathcal{I}$ and
$\mathcal{C}_i^{\,n+1} := \op{vec}(\vec{C}_i^{n+1}) \in \mathbb{R}^6$ for every node $\alpha\in\mathcal{J}$, leads to the usual nonlinear algebraic system
$\vec{R}_q^{\,n+1}(\vec{q}^{\,n+1},\mathcal{C}_i^{\,n+1}) = \vec{0}$.

\paragraph{Full-discrete evolution equation.}

Analogously, inserting the finite element approximations
$\vec{C}_{i,h}$ and $\delta\vec{\Sigma}_{h}$ into the midpoint weak
evolution equation \eqref{eq:weak_evol_midpoint_M} and using the
arbitrariness of the coefficients $\delta\vec{\Sigma}_\alpha$ yields
\begin{equation}
  \sum_{\alpha\in\mathcal{J}}
    \delta\vec{\Sigma}_\alpha :
    \vec{R}_{\mathrm{visc},\alpha}^{\,n+1}
  = 0,
\end{equation}
with viscous residuals
\begin{equation}\label{eq:viscousResidual}
  \vec{R}_{\mathrm{visc},\alpha}^{\,n+1}
  :=
  \int_{\mathcal{B}_0}
    H_\alpha(\vec{X}) 
    \Bigl(
      \frac{\vec{C}_{i,h}^{\,n+1} - \vec{C}_{i,h}^{\,n}}{h}
      - 2\,\vec{C}_{i,h}^{n+\frac12}\,\mathbb{V}^{-1} : \vec{M}_{h}^{n+\frac12,T}
    \Bigr)\,\mathrm{d}V.
\end{equation}
Collecting all $\vec{R}_{\mathrm{visc},\alpha}^{\,n+1}$ in a global vector $\vec{R}_{\mathrm{visc}}^{\,n+1} = \bigcup\limits_\alpha\op{vec}\left(\vec{R}_{\mathrm{visc},\alpha}^{\,n+1}\right)$ leads to the full-discrete evolution equation $\vec{R}_{\mathrm{visc}}^{\,n+1}(\vec{q}^{\,n+1},\mathcal{C}_i^{\,n+1}) = \vec{0}$. 
At this stage, the choice of the internal-variable shape functions
$H_\beta$ is still completely general. In the classical quadrature-point
based setting, one chooses a discontinuous approximation for the internal variable and
Dirac-type test functions for $\delta\vec{\Sigma}_h$.  Kronecker functions, as the discrete counterparts of Dirac pulses, are used
to represent the internal variable at the quadrature-points. With this choice, the viscous residuals reduce to the pointwise algebraic
equations
\[
\frac{\vec{C}_{i,h}^{\,n+1} - \vec{C}_{i,h}^{\,n}}{h}
      - 2\vec{C}_{i,h}^{n+\frac12}\,\mathbb{V}^{-1} : \vec{M}_{h}^{n+\frac12,T} = \vec{0},
\]
at each quadrature-point. This specialization will be
discussed later in the context of the monolithic Newton system and the
condensation of the internal variables.

\subsection{Role of the evolution law in the coupled problem}
\label{sec:evol_as_constraint_new}

Equations \eqref{eq:residual} and \eqref{eq:viscousResidual} 
 define the weak form of the coupled viscoelastic problem
on the time interval $[t_n,t_{n+1}]$. From the continuum viewpoint, \eqref{eq:weak_evol_Ci} is a constitutive
ODE for $\vec{C}_i$ enforced by thermodynamics. From the perspective of the
fully discrete nonlinear system, however, the relation
\eqref{eq:viscousResidual} appears as an additional set of algebraic
equations at time $t_{n+1}$.

In other words, the evolution law acts as an internal constraint in
the space of discrete configurations $(\vec{q}^{\,n+1},\mathcal{C}_i^{\,n+1})$.
This interpretation is crucial for the monolithic treatment of the coupled
problem: it allows us to assemble a
single Newton system for both primary variables and, subsequently, to condense
the internal variable $\vec{C}_i$ by Schur complements or null-space
projectors without ever performing a staggered update at the level of the
material law.
\section{Monolithic Newton system and condensation of the internal variable}\label{sec:solution}

\color{black}
Based on the midpoint-discrete weak forms of the momentum balance and
the evolution equation, we now assemble the nonlinear algebraic step
problem on the interval $[t_n,t_{n+1}]$. The unknowns are the endpoint
values collected in
\color{black}
\[
  z^{n+1} =
  \begin{bmatrix}
    q^{n+1}\\
    C_i^{n+1}
  \end{bmatrix}.
\]
The coupled residual is then
\[
 R^{n+1}(z^{n+1};z^n)
=
\begin{bmatrix}
  R_q^{n+1}(q^{n+1},C_i^{n+1};q^n,C_i^n)\\
  R_C^{n+1}(q^{n+1},C_i^{n+1};q^n,C_i^n)
\end{bmatrix}
=0,
\]
which has to be fulfilled up to the chosen Newton stopping criterion. In the following, we suppress the time index $n+1$ and, for brevity, the
subscript $i$ on the internal variable whenever possible. We denote the viscous residual by
$\vec{R}_{\mathrm{visc}} \equiv \vec{R}_\mathcal{C}$ and use the block notation
\[
  \vec{K}_{\alpha\beta}
  :=
  \frac{\partial \vec{R}_\alpha}{\partial \beta},\quad\alpha,\beta\in\{q,\mathcal{C}\}
\]
where the first index refers to the residual being linearized and the
second index to the variable. In particular,
$\vec{K}_{\mathcal{C}\mathcal{C}}$ denotes the linearization of
$\vec{R}_\mathcal{C}$ with respect to $\mathcal{C}$.

\color{black}
It is important to emphasize that the internal constraint manifold is a
time-discrete object. For fixed data at time \(t_n\), the equation
\(R_C^{n+1}(z^{n+1};z^n)=0\) defines the admissible endpoint states of the
chosen time-integration scheme on \([t_n,t_{n+1}]\). Thus, the superscript
\(n+1\) refers to the unknown endpoint state of the step problem, whereas
the constitutive quantities entering \(R_C^{n+1}\) are evaluated at the
midpoint state.
\color{black}

\subsection{Monolithic Newton system}

Linearizing the residual yields the block system
\begin{equation}
  \begin{bmatrix}
    \vec{K}_{qq} & \vec{K}_{q\mathcal{C}}\\[0.3em]
    \vec{K}_{\mathcal{C}q} & \vec{K}_{\mathcal{C}\mathcal{C}}
  \end{bmatrix}
  \begin{bmatrix}
    \Delta \vec{q} \\
    \Delta \mathcal{C}
  \end{bmatrix}
  =
  -
  \begin{bmatrix}
    \vec{R}_q \\
    \vec{R}_{\mathcal{C}}
  \end{bmatrix}.
  \label{eq:monolithic_block_system}
\end{equation}

The blocks have the following interpretations:

\begin{itemize}
  \item $\vec{K}_{qq}$ : stiffness-like tangent from equilibrium equation,
  \item $\vec{K}_{q\mathcal{C}}$ : sensitivity of internal forces to the internal variable,
  \item $\vec{K}_{\mathcal{C}q}$ : sensitivity of evolution equation to the deformation,
  \item $\vec{K}_{\mathcal{C}\mathcal{C}}$ : tangent of the evolution equation.
\end{itemize}

In general,
\[
  \vec{K}_{q\mathcal{C}} \neq \vec{K}_{\mathcal{C}q}\tp,
\]
since the evolution law for $\mathcal{C}_i$ is a constitutive ODE rather than a variational constraint; hence the monolithic tangent is non-symmetric.

\color{black}{
\subsection{Local solvability and convexity considerations}
\label{sec:local_solvability_convexity}

The block system \eqref{eq:monolithic_block_system} should not be interpreted, in general, as the Euler equation of a globally convex minimization problem in the combined variables $(\vec{q},\mathcal{C})$. Such a property cannot be inferred from the solution architecture alone. It depends on the particular choice of free energy, dissipation potential, time discretization, loading and admissible set of internal variables. Moreover, the evolution equation is introduced here as a constitutive internal constraint. The resulting monolithic tangent is therefore generally non-symmetric and is not, in general, the Hessian of a scalar incremental potential.

This observation does not affect the condensation procedure. The latter is a local algebraic reduction of the Newton linearization and requires only the local regularity of the internal block. More precisely, in a neighbourhood of a regular solution $(\vec{q}^{\ast},\mathcal{C}^{\ast})$ of
\[
  \vec{R}_{\mathcal{C}}(\vec{q}^{\ast},\mathcal{C}^{\ast}) = \vec{0},
\]
we assume that $\vec{K}_{\mathcal{C}\mathcal{C}}$ is nonsingular on the discrete internal-variable space. By the implicit function theorem, this implies that the internal equation locally defines an implicit mapping
\[
  \mathcal{C}^{\ast} = \mathcal{C}^{\ast}(\vec{q})
\]
with derivative
\begin{equation}
  \frac{\partial \mathcal{C}^{\ast}}{\partial \vec{q}}
  =
  -\vec{K}_{\mathcal{C}\mathcal{C}}^{-1}\vec{K}_{\mathcal{C}q}.
  \label{eq:implicit_derivative_Cq}
\end{equation}
This is exactly the regularity condition that is also required in the classical nested Gauss-point update: without an invertible local internal tangent, neither the local Newton solve nor the corresponding condensed algorithmic tangent is well defined. Thus, the monolithic formulation does not introduce an additional regularity assumption. Retaining the contribution associated with the non-vanishing internal residual
on the condensed right-hand side modifies the Newton correction away from the
constraint manifold, but does not alter the local regularity requirement.

For the constitutive model considered in this work, positive viscosity parameters and convexity of the local dissipation potential ensure thermodynamic admissibility and local dissipation. These material assumptions should, however, be distinguished from global convexity of the fully discrete algebraic residual. The present contribution therefore does not rely on a global convexity statement for $(\vec{q},\mathcal{C})$, but on the assumed local nonsingularity of $\vec{K}_{\mathcal{C}\mathcal{C}}$, which is common to both the monolithic and the classical condensed formulations.

\subsection{Locality and finite-element coupling of the internal evolution equation}
\label{sec:locality_FE_coupling}

The continuum evolution equation is local at each material point $X$; it contains no spatial derivatives of $\mathcal{C}_i$. The spatial finite-element discretization of its weak form nevertheless depends on the chosen approximation space for $\mathcal{C}_i$ and for the associated test functions. To make this point explicit, let
\[
  \mathcal{G}(\vec{q},\mathcal{C})
  :=
  2\,\mathcal{C}\,\bigl(\vec{V}^{-1}:\vec{M}(\vec{q},\mathcal{C})\tp\bigr)
\]
denote the right-hand side of the midpoint-discrete evolution equation. On an element $B_0^e$, the discrete internal residual has the generic form
\begin{equation}
  \vec{R}_{\mathcal{C},\alpha}^{e}
  =
  \int_{B_0^e}
    H_\alpha
    \left(
      \sum_{\beta\in J^e} H_\beta\,\frac{\mathcal{C}^{n+1}_{\beta}-\mathcal{C}^{n}_{\beta}}{h}
      -
      \mathcal{G}\!\left(\vec{q}^{\,n+1},
      \sum_{\beta\in J^e} H_\beta\,\mathcal{C}^{n+\frac12}_{\beta}
      \right)
    \right)
  \,\mathrm{d}V .
  \label{eq:element_internal_residual_generic}
\end{equation}
Consequently, even a local evolution equation leads to an element-level algebraic coupling between all internal degrees of freedom whose basis functions overlap. In the simplest linear part, this coupling is the usual mass-like matrix
\[
  \vec{M}^{e}_{\alpha\beta}
  =
  \int_{B_0^e} H_\alpha H_\beta\,\mathrm{d}V,
\]
and the corresponding internal tangent has the structure
\begin{equation}
  \vec{K}_{\mathcal{C}\mathcal{C},\alpha\beta}^{e}
  =
  \int_{B_0^e}
    H_\alpha
    \left[
      \frac{1}{h}H_\beta\,\mathbb{I}
      -
      \partial_{\mathcal{C}}\mathcal{G}(\vec{q},\mathcal{C}_{h})
      \bigl[H_\beta\,\mathbb{I}\bigr]
    \right]
  \,\mathrm{d}V .
  \label{eq:element_KCC_generic}
\end{equation}
This coupling is, however, not a diffusion-type regularization. No gradients of $\mathcal{C}_i$ or of the test functions occur in \eqref{eq:element_internal_residual_generic}--\eqref{eq:element_KCC_generic}; hence the formulation introduces neither a Laplace-type operator nor an internal length scale or flux across element boundaries. The coupling is solely the consequence of a Galerkin projection of a pointwise evolution equation onto the chosen finite-dimensional internal-variable space.

The distinction is important for the block structure. Since no inter-element continuity is imposed for $\mathcal{C}_i$, polynomial choices of $H_\beta$ lead at most to elementwise dense blocks,
\[
  \vec{K}_{\mathcal{C}\mathcal{C}}
  =
  \bigoplus_{e}\vec{K}_{\mathcal{C}\mathcal{C}}^{e},
\]
but not to a global coupling of the internal variables between neighbouring elements. In the quadrature-point-localized limit, where the basis is chosen in the sense of Dirac or Kronecker functions satisfying $H_\alpha(X_g)=\delta_{\alpha g}$ at the quadrature points, the element block further decomposes into independent pointwise contributions,
\[
  \vec{K}_{\mathcal{C}\mathcal{C}}^{e}
  =
  \bigoplus_{g}\vec{K}_{\mathcal{C}\mathcal{C}}^{e,g},
\]
and the standard local Gauss-point update is recovered. Thus, the classical Dirac-type discretization is a particular member of the same discrete framework, whereas piecewise constant, piecewise linear or higher-order discontinuous internal-variable spaces correspond to different local projections of the same evolution equation.

The influence of this choice is therefore a discretization effect: on coarse meshes, different internal-variable spaces may produce slightly different projected internal fields and stresses, whereas under refinement they should approach the same local evolution problem, provided the approximation is stable and sufficiently resolved. This effect is assessed numerically below by comparing Dirac-type, elementwise constant and polynomial internal-variable approximations.
}

\color{black}{

\subsection{Condensation via the Schur complement}\label{sec:staggered_solution}

The block system \eqref{eq:monolithic_block_system} can be solved for $\Delta \mathcal{C}$. In particular, from the second row we obtain
\begin{equation}
  \Delta \mathcal{C}
  =
  -\vec{K}_{\mathcal{C}\mathcal{C}}^{-1} \vec{R}_\mathcal{C}
   -\vec{K}_{\mathcal{C}\mathcal{C}}^{-1} \vec{K}_{\mathcal{C}q}\,\Delta \vec{q}.
  \label{eq:Delta_c_solution}
\end{equation}
The inverse of $\vec{K}_{\mathcal{C}\mathcal{C}}$ is understood in the sense of the local or elementwise internal solve discussed in Sections~\ref{sec:local_solvability_convexity} and \ref{sec:locality_FE_coupling}. For quadrature-point-localized internal variables this block is diagonal over the integration points; for discontinuous polynomial internal-variable spaces it is block diagonal over the elements.

Insertion into the first row yields\begin{align}
  \vec{K}_{qq}\,\Delta \vec{q}
  + \vec{K}_{q\mathcal{C}}\Bigl(
      -\vec{K}_{\mathcal{C}\mathcal{C}}^{-1} \vec{R}_\mathcal{C}
      -\vec{K}_{\mathcal{C}\mathcal{C}}^{-1} \vec{K}_{\mathcal{C}q}\,\Delta \vec{q}
    \Bigr)
  &= -\vec{R}_q.
\end{align}
Collecting the terms in $\Delta \vec{q}$ yields the condensed system
\begin{equation}
  \underbrace{
  \Bigl(
    \vec{K}_{qq}
    - \vec{K}_{q\mathcal{C}} \vec{K}_{\mathcal{C}\mathcal{C}}^{-1} \vec{K}_{\mathcal{C}q}
  \Bigr)
  }_{\displaystyle \vec{K}_{qq}^\mathrm{cond}}
  \Delta \vec{q}
  =
  -
  \Bigl(
    \vec{R}_q
    - \vec{K}_{q\mathcal{C}} \vec{K}_{\mathcal{C}\mathcal{C}}^{-1} \vec{R}_\mathcal{C}
  \Bigr).
  \label{eq:condensed_Schur}
\end{equation}
\color{black}
The matrix $\vec{K}_{qq}^\mathrm{cond}$ is the Schur complement of the internal variable block. Once $\Delta \vec{q}$ is computed, $\Delta \mathcal{C}$ is recovered from \eqref{eq:Delta_c_solution}.
In an element-level implementation, we introduce
\begin{equation}
 A_e=K_{CC}^{-1}K_{Cq},
 \qquad
 b_e=K_{CC}^{-1}R_C.
\end{equation}
Both quantities can be obtained in a single local linear solve with multiple
right-hand sides and retained for the subsequent reconstruction
of the internal increment. This avoids storing the factorization of \(K_{CC}\)
beyond the local condensation step.
\color{black}

\paragraph{Classical (nested) Gauss-point condensation.}
In the classical quadrature-point setting, the internal variable is eliminated by enforcing the
internal equation locally (and nonlinearly) at every integration point. More precisely, for a given
iterate $\vec{q}^{(k)}$ one solves
\begin{equation}
  \vec{R}_{\mathcal{C}}\bigl(\vec{q}^{(k)}, \mathcal{C}\bigr) = \vec{0}
\end{equation}
by a local Newton method at each Gauss point up to an internal tolerance $\epsilon_{\mathrm{int}}$.
This yields an (implicit) solution $\mathcal{C}^\star(\vec{q})$ and hence a reduced equilibrium
residual $\widehat{\vec{R}}_q(\vec{q}) := \vec{R}_q\!\left(\vec{q}, \mathcal{C}^\star(\vec{q})\right)$.
This leads to the condensed tangent
\begin{equation}
  \widehat{\vec{K}}_{qq}
  =
  \frac{\partial \widehat{\vec{R}}_q}{\partial \vec{q}}
  =
  \vec{K}_{qq}
  -
  \vec{K}_{q\mathcal{C}}\,\vec{K}_{\mathcal{C}\mathcal{C}}^{-1}\,\vec{K}_{\mathcal{C}q},
\end{equation}
which coincides with the Schur complement $\vec{K}_{qq}^{\mathrm{cond}}$ in \eqref{eq:condensed_Schur}.
However, because the classical scheme enforces $\vec{R}_{\mathcal{C}}=\vec{0}$ before assembling the
global step, the right-hand side reduces to $-\vec{R}_q$ and \eqref{eq:condensed_Schur} becomes
\begin{equation}
  \vec{K}_{qq}^{\mathrm{cond}}\,\Delta\vec{q} = -\vec{R}_q.
\end{equation}
In contrast, the monolithic condensation retains the coupling term
$-\vec{K}_{q\mathcal{C}}\vec{K}_{\mathcal{C}\mathcal{C}}^{-1}\vec{R}_{\mathcal{C}}$, which accounts for
non-vanishing internal residuals during the global Newton iterations and avoids a nested local
Newton solve at each quadrature point.

\subsection{Condensation via a null-space operator}

The Schur complement \eqref{eq:condensed_Schur} can also be written in operator form, which provides a geometric interpretation in terms of the tangent space of the internal constraint.

\paragraph{Row operator and Schur form.}

We first introduce a row condensation operator that eliminates the internal variable from the linear system. Define
\[
  \vec{P}_\mathrm{row}
  :=
  \begin{bmatrix}
    \vec{I} & -\,\vec{K}_{q\mathcal{C}}\,\vec{K}_{\mathcal{C}\mathcal{C}}^{-1}
  \end{bmatrix}.
\]
A short block computation shows that
\[
  \vec{P}_\mathrm{row}\,
  \vec{K}
  =
  \begin{bmatrix}
    \vec{K}_{qq} - \vec{K}_{q\mathcal{C}}\vec{K}_{\mathcal{C}\mathcal{C}}^{-1}\vec{K}_{\mathcal{C}q}
    &
    \vec{0}
  \end{bmatrix}.
\]
Multiplying \eqref{eq:monolithic_block_system} from the left with $\vec{P}_\mathrm{row}$ gives
\[
  \vec{P}_\mathrm{row}\,\vec{K}\,\Delta \vec{z}
  =
  -\vec{P}_\mathrm{row}\,\vec{R},
\qquad
  \vec{R} =
  \begin{bmatrix}
    \vec{R}_q \\ \vec{R}_\mathcal{C}
  \end{bmatrix},
\]
such that we obtain
\[
  \bigl(
    \vec{K}_{qq}
    - \vec{K}_{q\mathcal{C}} \vec{K}_{\mathcal{C}\mathcal{C}}^{-1} \vec{K}_{\mathcal{C}q}
  \bigr)\Delta \vec{q}
  =
  -\bigl(
    \vec{R}_q
    - \vec{K}_{q\mathcal{C}} \vec{K}_{\mathcal{C}\mathcal{C}}^{-1} \vec{R}_\mathcal{C}
  \bigr),
\]
which is exactly the condensed system \eqref{eq:condensed_Schur}.

\paragraph{Null-space basis and tangent space.}

Next, we regard the linearized evolution equation as a constraint for the admissible Newton increments. The second block row of \eqref{eq:monolithic_block_system} reads
\[
  \vec{R}_\mathcal{C} + \vec{K}_{\mathcal{C}q}\,\Delta \vec{q} + \vec{K}_{\mathcal{C}\mathcal{C}}\,\Delta \mathcal{C} = \vec{0},
\]
and can be written in abstract form as
\[
  \vec{R}_\mathcal{C} + \vec{L}\,\Delta \vec{z} = \vec{0},
  \quad\text{where}\quad
  \vec{L} :=
  \begin{bmatrix}
    \vec{K}_{\mathcal{C}q} & \vec{K}_{\mathcal{C}\mathcal{C}}
  \end{bmatrix}.
\]
At a converged solution we have $\vec{R}_\mathcal{C}=\vec{0}$, and the tangent space of the constraint manifold is characterized by the homogeneous equation
$\vec{L}\,\Delta \vec{z} = \vec{0}$. A basis of admissible increments is provided by the column operator
\[
  \vec{P}_\mathrm{col}
  :=
  \begin{bmatrix}
    \vec{I} \\
    -\,\vec{K}_{\mathcal{C}\mathcal{C}}^{-1}\,\vec{K}_{\mathcal{C}q}
  \end{bmatrix},
  \qquad
  \Delta \vec{z} = \vec{P}_\mathrm{col}\,\Delta \vec{q},
\]
since
\[
  \vec{L}\,\vec{P}_\mathrm{col}
  =
  \begin{bmatrix}
    \vec{K}_{\mathcal{C}q} & \vec{K}_{\mathcal{C}\mathcal{C}}
  \end{bmatrix}
  \begin{bmatrix}
    \vec{I} \\
    -\,\vec{K}_{\mathcal{C}\mathcal{C}}^{-1}\,\vec{K}_{\mathcal{C}q}
  \end{bmatrix}
  =
  \vec{K}_{\mathcal{C}q} - \vec{K}_{\mathcal{C}\mathcal{C}}\vec{K}_{\mathcal{C}\mathcal{C}}^{-1}\vec{K}_{\mathcal{C}q}
  =
  \vec{0}.
\]
Away from the internal constraint manifold, the admissible Newton
correction is affine rather than linear:
\begin{equation}
  \Delta\vec{z}
  =
  \begin{bmatrix}
    \vec{0}\\
    -\vec{K}_{\mathcal{C}\mathcal{C}}^{-1}\vec{R}_{\mathcal{C}}
  \end{bmatrix}
  +
  \vec{P}_\mathrm{col}\,\Delta\vec{q} .
\end{equation}
At an internally converged state, $\vec{R}_{\mathcal{C}}=\vec{0}$, this affine space
reduces to the tangent space parameterized by $\vec{P}_\mathrm{col}$.
Restricting the Newton increment to this tangent space and testing the linear system
with the row operator yields the reduced system
\begin{equation}
  \vec{P}_\mathrm{row}\,\vec{K}\,\vec{P}_\mathrm{col}\,\Delta \vec{q}
  =
  -\,\vec{P}_\mathrm{row}\,\vec{R}.
  \label{eq:nullspace_reduced}
\end{equation}
Using the block structure of $\vec{K}$ and $\vec{R}$, one readily verifies
\[
  \vec{P}_\mathrm{row}\,\vec{K}\,\vec{P}_\mathrm{col}
  =
  \vec{K}_{qq} - \vec{K}_{q\mathcal{C}}\,\vec{K}_{\mathcal{C}\mathcal{C}}^{-1}\,\vec{K}_{\mathcal{C}q},
  \qquad
  \vec{P}_\mathrm{row}\,\vec{R}
  =
  \vec{R}_q - \vec{K}_{q\mathcal{C}}\,\vec{K}_{\mathcal{C}\mathcal{C}}^{-1}\,\vec{R}_\mathcal{C},
\]
so that \eqref{eq:nullspace_reduced} is identical to the Schur complement system \eqref{eq:condensed_Schur}. 
Since the monolithic tangent is generally non-symmetric, the left and right operator differ:
$\vec{P}_\mathrm{col}$ parameterizes admissible increments in the tangent space
of the internal constraint, while $\vec{P}_\mathrm{row}$ selects those test
directions in which the internal variables have already been condensed.

\paragraph{Geometric interpretation of the null-space basis.}

The set of all configurations that satisfy the internal equation defines the internal constraint manifold
\[
  \mathfrak{M}
  := \bigl\{
        \vec{z} \in \mathbb{R}^{n_q+n_\mathcal{C}}
        \,\big|\,
        \vec{R}_\mathcal{C}(\vec{z}) = \vec{0}
     \bigr\}.
\]

Consider a configuration $\vec{z}^\ast = (\vec{q}^\ast,\mathcal{C}^\ast)$ on
$\mathfrak{M}$, i.e.\ $\vec{R}_\mathcal{C}(\vec{z}^\ast) = \vec{0}$. A perturbation
$\Delta\vec{z}$ is admissible in first order if it keeps the constraint
satisfied up to quadratic terms,
\[
  \vec{R}_\mathcal{C}(\vec{z}^\ast + \Delta\vec{z})
  = \mathcal{O}(\|\Delta\vec{z}\|^2).
\]
Linearizing $\vec{R}_\mathcal{C}$ at $\vec{z}^\ast$ yields
\[
  \vec{R}_\mathcal{C}(\vec{z}^\ast + \Delta\vec{z})
  \approx
  \vec{R}_\mathcal{C}(\vec{z}^\ast)
  + \underbrace{
      \frac{\partial\vec{R}_\mathcal{C}}{\partial\vec{z}}(\vec{z}^\ast)
    }_{=[\,\vec{K}_{\mathcal{C}q}\ \ \vec{K}_{\mathcal{C}\mathcal{C}}\,]}
    \Delta\vec{z}.
\]
Since $\vec{R}_\mathcal{C}(\vec{z}^\ast)=\vec{0}$, admissible increments must satisfy the linearized constraint
\begin{equation}
  \vec{K}_{\mathcal{C}q}\,\Delta\vec{q}
  + \vec{K}_{\mathcal{C}\mathcal{C}}\,\Delta\vec{\mathcal{C}}
  = \vec{0}.
  \label{eq:linearized_internal_constraint}
\end{equation}
The tangent space of $\mathfrak{M}$ at $\vec{z}^\ast$ is therefore the null-space of the block row $[\,\vec{K}_{\mathcal{C}q}\ \ \vec{K}_{\mathcal{C}\mathcal{C}}\,]$,
\[
  T_{\vec{z}^\ast}\mathfrak{M}
  =
  \bigl\{
    \Delta\vec{z}
    \,\big|\,
    \vec{K}_{\mathcal{C}q}\,\Delta\vec{q}
    + \vec{K}_{\mathcal{C}\mathcal{C}}\,\Delta\mathcal{C} = \vec{0}
  \bigr\}.
\]
Assuming that $\vec{K}_{\mathcal{C}\mathcal{C}}$ is nonsingular in the considered regime - which is the same local regularity condition required by the classical local constitutive update - the linearized constraint can be solved for $\Delta\mathcal{C}$ in terms of $\Delta\vec{q}$ as
\[
  \Delta\mathcal{C}
  = -\,\vec{K}_{\mathcal{C}\mathcal{C}}^{-1}\,\vec{K}_{\mathcal{C}q}\,\Delta\vec{q},
\]
as shown above. The columns of the matrix $\vec{P}_\mathrm{col}$ therefore form a basis of the
null-space of the linearized constraint, and we have the geometric identity
\[
  \operatorname{Im}(\vec{P}_\mathrm{col})
  =
  \ker
  \begin{bmatrix}
    \vec{K}_{\mathcal{C}q} & \vec{K}_{\mathcal{C}\mathcal{C}}
  \end{bmatrix}
  =
  T_{\vec{z}^\ast}\mathfrak{M}.
\]
In this sense, $\vec{P}_\mathrm{col}$ is a null-space basis that spans
the tangent space of the internal constraint manifold. The actual reduction of
the Newton system is then performed by restricting the increments to this
tangent space via $\Delta\vec{z} = \vec{P}_\mathrm{col}\,\Delta\vec{q}$ and
premultiplying the monolithic system with the row operator
$\vec{P}_\mathrm{row}$. The null-space formulation provides a clean geometric
interpretation of the Schur complement condensation: Newton steps are confined to
the tangent space of the internal constraint manifold, and the condensed operator
$\vec{K}_{qq}^\mathrm{cond}$ is precisely the restriction of $\vec{K}$ to this space.}
\section{Numerical examples}\label{sec:examples}

In this section, we compare the proposed monolithic Newton strategy, in which the internal variable is condensed at the level of the global linearized system, with a classical nested Gauss-point condensation scheme, where the internal evolution equation is solved locally at each quadrature point by an inner Newton loop (see Section~\ref{sec:solution}). Unless stated otherwise, both approaches employ identical spatial discretizations and linear solver settings; any observed differences therefore reflect the nonlinear solution strategy.

Throughout the examples, we employ the finite viscoelastic Simo--Taylor material introduced in Section~\ref{sec:energy}. The equilibrium/non-equilibrium split of the free energy density reads
\[
  \Psi(\vec{C},\mathcal{C}_i)
  = \Psi_{\mathrm{eq}}(\vec{C}) + \Psi_{\mathrm{neq}}(\vec{C}_e),
  \qquad \vec{C}_e = \vec{C}\,\mathcal{C}_i^{-1}.
\]
In our implementation, the equilibrium part is chosen as
\[
  \Psi_{\mathrm{eq}}(\vec{C})
  =
  \frac{\mu}{2}\Bigl(\mathrm{tr}(\vec{C}) - d - 2\log(J)\Bigr)
  + \frac{\lambda}{2}\Bigl((\log(J))^2 + (J-1)^2\Bigr),
  \qquad J = \sqrt{\det(\vec{C})},
\]
and the non-equilibrium part is of the same form, evaluated at \(\vec{C}_e\),
\[
  \Psi_{\mathrm{neq}}(\vec{C}_e)
  =
  \frac{\mu_{\mathrm{visc}}}{2}\Bigl(\mathrm{tr}(\vec{C}_e) - d - 2\log(J_e)\Bigr)
  + \frac{\lambda_{\mathrm{visc}}}{2}\Bigl((\log(J_e))^2 + (J_e-1)^2\Bigr),
  \qquad J_e = \sqrt{\det(\vec{C}_e)}.
\]
Here, \(d\) denotes the dimension of the tensor representation used in the constitutive evaluation (in the present examples, \(d=3\) for both plane strain and 3D). The viscous flow is governed by the isotropic viscosity tensor
\[
  \mathbb{V} = V_{\mathrm{vol}}\,\mathbb{P}_{\mathrm{vol}} + 2V_{\mathrm{dev}}\,\mathbb{P}_{\mathrm{dev}},
\]
with parameters listed in Table~\ref{material_params} in terms of area, i.e.\ scaled by the (unit) thickness.

\begin{table}[h!]
	\centering
	\begin{tabular}{cc r @{ } l}
		\hline
		Parameter & Symbol & Value & Unit \\
		\hline
		Density                  & $\rho_0$   & 10 & [kg/m²] \\ 
		First Lam$\'e$ constant  & $\lambda$  & 30000 & [J/m²]   \\
		Second Lam$\'e$ constant  & $\mu$      & 7500 & [J/m²]\\
		Deviatoric viscosity     & $V_{dev}$  & 10000 & [Js/m²]  \\
		Volumetric viscosity     & $V_{vol}$  & 50000 & [Js/m²]  \\
		Viscous First Lam$\'e$ constant & $\lambda_{visc}$  & 30000 & [J/m²]  \\
		Viscous Second Lam$\'e$ constant & $\mu_{visc}$  & 7500 & [J/m²] \\
		\hline
	\end{tabular}
	\caption{Material parameters. Note that we have calibrated the set of parameters per area.}
	\label{material_params}
\end{table}


For verification purposes, three algebraically related implementations are
available. The fully monolithic formulation solves the complete coupled Newton
system in the displacement and internal-variable increments. The proposed
monolithically condensed formulation eliminates the internal increment locally
at the linearized level and reconstructs it after the reduced global solve.
The classical nested formulation first solves the nonlinear internal equations
locally and subsequently assembles the reduced equilibrium system. Unless stated
otherwise, the numerical comparisons below refer to the proposed monolithically
condensed and the classical nested formulations; the fully monolithic variant is
used only as an algebraic reference.

All condensed linear systems are solved with MATLAB's sparse direct solver
through the backslash operator (\texttt{mldivide}). The generally
non-symmetric tangent matrices are passed to the solver without additional
symmetrization. The nonlinear equations are solved by a damped Deuflhardt
Newton method. For the classical nested formulation, the local constitutive
Newton tolerance is specified separately. All reported wall-clock times were
measured on Intel Xeon Gold 6148 CPU and 756 GB RAM using MATLAB 2025b with a pre-initialized parallel
pool of 20 workers; pool startup and postprocessing were excluded.

\subsection{Algebraic equivalence for the 2D Cook's membrane}
\label{sec:example2D}

We first verify that the three solution strategies introduced in
Section~\ref{sec:solution} produce the same discrete response. To this end, we
consider the two-dimensional Cook's membrane benchmark on the reference domain
defined by the corner points
\[
  \vec{P}_1=(0,0),\qquad
  \vec{P}_2=(480,440),\qquad
  \vec{P}_3=(480,600),\qquad
  \vec{P}_4=(0,440),
\]
where all coordinates are given in \([\mathrm{mm}]\). The initial velocity is
zero. A plane-strain setting is adopted by embedding the in-plane deformation
gradient into three dimensions,
\[
  \vec{F} =
  \begin{bmatrix}
    \vec{F}_{2D} & \vec{0}\\
    \vec{0} & 1
  \end{bmatrix}.
\]
With this choice, the standard three-dimensional kinematic relations reduce
consistently to plane strain. Details of the corresponding reduction are given
in \cite{hesch_framework_2017}.

The left edge is fixed by homogeneous Dirichlet boundary conditions,
\[
  \Gamma^\varphi
  :=
  \left\{
    \vec{X}\in\mathbb{R}^2
    \,\middle|\,
    X_1=0,\;0<X_2<440
  \right\}.
\]
On the right edge,
\[
  \Gamma_r^\sigma
  :=
  \left\{
    \vec{X}\in\mathbb{R}^2
    \,\middle|\,
    X_1=480,\;440<X_2<600
  \right\},
\]
we prescribe the time-dependent surface traction
\[
  \vec{T}(\vec{X},t)
  =
  \Lambda(t)\,p\,\bar{\vec{T}},
  \qquad
  \bar{\vec{T}}=(-750,1000),
\]
with
\[
  \Lambda(t)
  =
  \begin{cases}
    t/T_{\mathrm{end}}, & 0\leq t\leq T_{\mathrm{end}},\\
    1,                  & t>T_{\mathrm{end}},
  \end{cases}
\]
where \(p\) is a dimensionless load multiplier. Body forces are neglected.

The deformation is discretized by \(20\times20\) bilinear Q1 elements with four
Gauss points per element. The internal variable \(\mathcal{C}_i\) is stored at
the quadrature points, as described in Section~\ref{sec:spatial_FE}. The time
interval \(T_{\mathrm{end}}=10\,\mathrm{s}\) is divided into ten equal time
steps. This deliberately moderate discretization serves only as an algebraic
solver-comparison benchmark; spatial and temporal convergence are investigated
separately in Section~\ref{sec:conv_study}.

Three implementations are compared:

\begin{enumerate}
  \item the fully monolithic formulation, in which displacement and internal
        variables are retained in the global Newton system;
  \item the proposed monolithically condensed formulation, in which the
        internal-variable increment is eliminated locally from the linearized
        system and reconstructed after the reduced global solve;
  \item the classical nested formulation, in which the nonlinear constitutive
        problem is solved locally at every quadrature point before assembling
        the reduced global equilibrium problem.
\end{enumerate}

For the present mesh, the fully monolithic formulation contains \(7282\)
globally coupled unknowns, whereas both condensed formulations require only
\(882\) global displacement unknowns. The same \(6400\) internal-variable
components are stored in all implementations; the distinction is whether they
enter the global linear system.

\begin{figure}[tb]
  \centering
  \includegraphics[width=\textwidth]{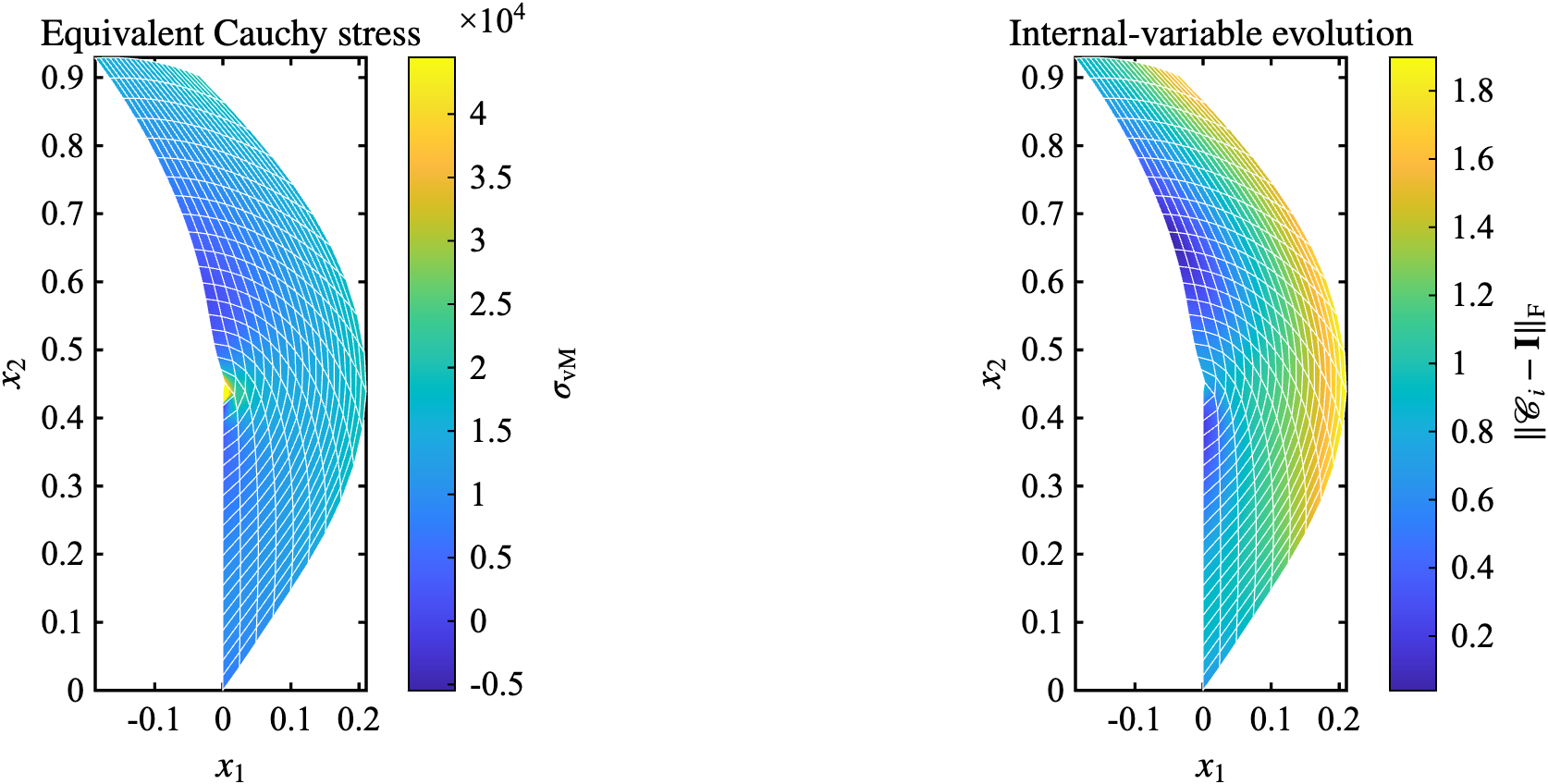}
  \caption{Final state of the 2D Cook's membrane for \(p=4\), computed with
  the monolithically condensed formulation. Left: equivalent Cauchy stress
  \(\sigma_{\mathrm{vM}}\). Right: internal-variable measure
  \(\|\mathcal{C}_i-\mathbf{I}\|_{\mathrm F}\). The deformed configuration is
  plotted with unit scale. The upper limit of the stress color range is capped
  at \(35\,\%\) of the maximum projected nodal value in order to reveal the
  stress distribution away from the localized corner peak.}
  \label{fig:sec51_fields}
\end{figure}

Figure~\ref{fig:sec51_fields} illustrates the final state for the largest load
level \(p=4\), obtained with the monolithically condensed formulation. The
deformed configuration is shown without artificial amplification. The large
change of shape therefore reflects the actual finite-deformation response of
the benchmark. The left panel displays the equivalent Cauchy stress
\(\sigma_{\mathrm{vM}}\), whereas the right panel shows the scalar
internal-variable measure
\[
  \|\mathcal{C}_i-\mathbf{I}\|_{\mathrm F}.
\]
The stress field contains the expected localized peak at the upper corner of
the clamped boundary. To make the stress distribution in the remaining domain
visible, the upper limit of the displayed color range is restricted to
\(35\,\%\) of the maximum projected nodal value. The actual peak is therefore
outside the displayed color range.

The quantitative results are summarized in
Table~\ref{tab:sec51_solver_comparison}. The outer Newton iteration counts are
very similar for all formulations. Hence, the condensation does not alter the
nonlinear solution path in any relevant way. At the same time, the proposed
formulation reduces the dimension of the globally coupled system from \(7282\)
to \(882\) unknowns.

\begin{table}[tb]
  \centering
  \caption{Solver comparison for the 2D Cook's membrane. Here,
  \(N_{\mathrm{glob}}\) denotes the number of unknowns in the global linear
  system, \(\bar n_{\mathrm{N}}\) the average number of outer Newton
  iterations per time step, and \(n_{\mathrm{N}}^{\max}\) the maximum number
  of outer Newton iterations.}
  \label{tab:sec51_solver_comparison}
  \begin{tabular}{c l r c c c}
    \hline
    \(p\) & Formulation
          & \(N_{\mathrm{glob}}\)
          & \(\bar n_{\mathrm{N}}\)
          & \(n_{\mathrm{N}}^{\max}\)
          & \(\|u_{\mathrm{tip}}\|\) \\
    \hline
    1 & Fully monolithic          & 7282 & 4.7 & 5 & 0.391699 \\
      & Monolithically condensed  &  882 & 4.7 & 5 & 0.391699 \\
      & Classical nested          &  882 & 4.3 & 5 & 0.391699 \\
    \hline
    2 & Fully monolithic          & 7282 & 4.9 & 6 & 0.565657 \\
      & Monolithically condensed  &  882 & 4.9 & 6 & 0.565657 \\
      & Classical nested          &  882 & 4.5 & 5 & 0.565657 \\
    \hline
    4 & Fully monolithic          & 7282 & 5.4 & 7 & 0.743842 \\
      & Monolithically condensed  &  882 & 5.2 & 6 & 0.743842 \\
      & Classical nested          &  882 & 4.9 & 6 & 0.743842 \\
    \hline
  \end{tabular}
\end{table}

To assess the agreement beyond a single output quantity, the complete
displacement and internal-variable fields are compared with the fully
monolithic reference. Table~\ref{tab:sec51_field_errors} reports
\[
  e_\varphi
  =
  \frac{\|\varphi-\varphi_{\mathrm{mono}}\|}
       {\|\varphi_{\mathrm{mono}}\|},
  \qquad
  e_{\mathcal{C}}
  =
  \frac{\|\mathcal{C}_i-\mathcal{C}_{i,\mathrm{mono}}\|}
       {\|\mathcal{C}_{i,\mathrm{mono}}\|},
\]
together with the difference in the final tip displacement. The
monolithically condensed formulation agrees with the fully monolithic solution
essentially up to machine precision for \(p=1\) and \(p=2\). Even for the
largest load level, the relative displacement-field difference remains below
\(10^{-12}\). The classical nested formulation agrees within the prescribed
nonlinear solution accuracy; its slightly larger differences arise from the
separate termination of the local constitutive Newton iterations.

\begin{table}[tb]
  \centering
  \caption{Differences with respect to the fully monolithic reference solution.
  The norms include the complete discrete displacement and internal-variable
  fields.}
  \label{tab:sec51_field_errors}
  \begin{tabular}{c l c c c}
    \hline
    \(p\) & Formulation
          & \(\|\Delta u_{\mathrm{tip}}\|\)
          & \(e_\varphi\)
          & \(e_{\mathcal{C}}\) \\
    \hline
    1 & Monolithically condensed
      & \(3.27\times10^{-15}\)
      & \(1.46\times10^{-15}\)
      & \(4.94\times10^{-15}\) \\
      & Classical nested
      & \(3.69\times10^{-11}\)
      & \(1.95\times10^{-11}\)
      & \(4.27\times10^{-12}\) \\
    \hline
    2 & Monolithically condensed
      & \(2.91\times10^{-15}\)
      & \(1.44\times10^{-15}\)
      & \(3.75\times10^{-15}\) \\
      & Classical nested
      & \(1.07\times10^{-11}\)
      & \(4.90\times10^{-12}\)
      & \(8.43\times10^{-13}\) \\
    \hline
    4 & Monolithically condensed
      & \(2.07\times10^{-12}\)
      & \(8.72\times10^{-13}\)
      & \(2.06\times10^{-14}\) \\
      & Classical nested
      & \(5.17\times10^{-11}\)
      & \(1.86\times10^{-11}\)
      & \(1.66\times10^{-12}\) \\
    \hline
  \end{tabular}
\end{table}

\begin{figure}[tb]
  \centering
  \includegraphics[width=0.78\textwidth]{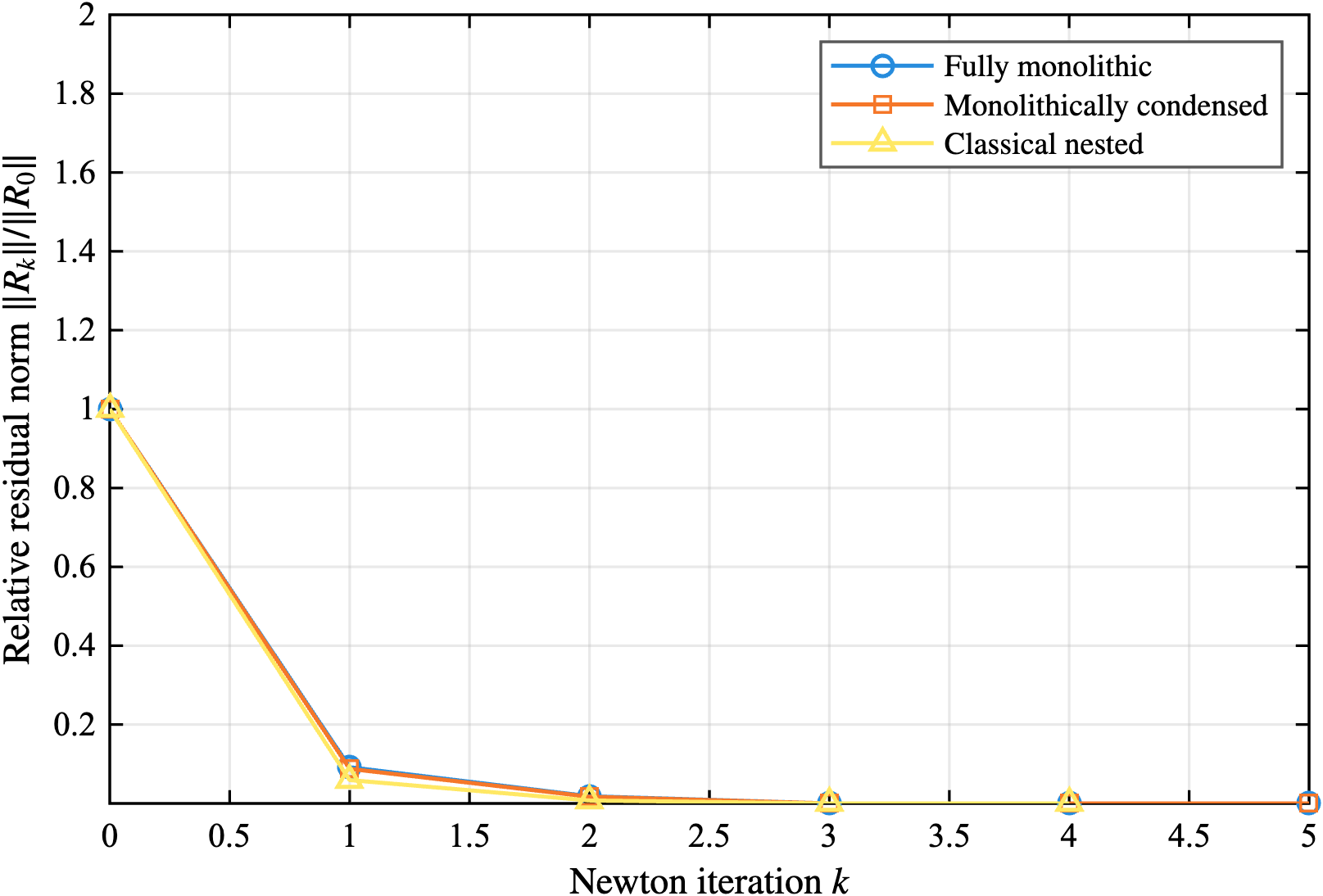}
  \caption{Normalized outer Newton residual in the final time step for
  \(p=4\). Each residual is normalized by the corresponding initial residual
  of the same formulation.}
  \label{fig:sec51_newton}
\end{figure}

Figure~\ref{fig:sec51_newton} shows the normalized residual history in the final
time step for \(p=4\). Since the fully monolithic and condensed residual vectors
have different dimensions and scalings, each curve is normalized by its own
initial residual. The resulting histories exhibit the same rapid local
convergence. In particular, static condensation does not degrade the
convergence characteristics of the outer Newton method.

Finally, the equivalence of the two reduced formulations is checked in a
ramp-and-hold test. The load with \(p=1\) is ramped over
\(0\leq t\leq10\,\mathrm{s}\) and then held constant until
\(t=15\,\mathrm{s}\). The monolithically condensed and classical nested
formulations yield the same final tip displacement,
\[
  u_{\mathrm{tip}}(15\,\mathrm{s})
  =
  (-0.355119,\;0.215424),
\]
with a difference of \(5.97\times10^{-11}\). Their average numbers of outer
Newton iterations per time step are \(4.27\) and \(3.93\), respectively.
Consequently, the agreement is retained over a loading history involving a
subsequent hold period.

The results of this section establish that the proposed condensation is an
algebraic reformulation of the fully monolithic Newton step: it preserves the
discrete solution and the outer Newton convergence while substantially reducing
the dimension of the global linear system. Its computational performance is
examined separately in the following section.

\subsection{Computational performance}\label{sec:performance}

Having demonstrated in the previous section that the proposed monolithically
condensed formulation reproduces the reference solution up to machine
precision, we now investigate its computational performance. The objective is
to quantify the reduction in wall-clock time that can be achieved by
eliminating the internal variables on the element level before assembling the
global nonlinear system.

Two complementary studies are performed. First, a mesh refinement study is
carried out for the two-dimensional Cook membrane using quadratic
displacement interpolation under a fixed load level of $p=0.5$. This study
investigates how the runtime develops with increasing problem size. Second, a
parameter study with increasing load levels is performed on a fixed mesh to
assess the influence of the nonlinear constitutive response on the overall
computational cost.

The reported wall-clock times include the complete nonlinear solution
procedure, including element assembly, local condensation, sparse
factorization and all Newton iterations.

\subsubsection{Scaling with problem size}

Table~\ref{tab:sec52_scaling} summarizes the measured runtimes for the
classical condensed formulation and the proposed monolithically condensed
formulation. Both approaches compute identical displacement fields and differ
only in the treatment of the internal variables during the global solution
process.

\begin{table}[tb]
  \centering
  \caption{Wall-clock times for the Q2 scaling study of the Cook membrane
  with load multiplier \(p=0.5\) and \(40\) time steps. For repeated
  computations, the median runtime is reported. The runtime ratio is defined
  as \(T_{\mathrm{nested}}/T_{\mathrm{monocond}}\).}
  \label{tab:sec52_scaling}
  \begin{tabular}{r r r r r}
    \hline
    Elements &
    \(N_u\) &
    \(T_{\mathrm{monocond}}\,[\mathrm{s}]\) &
    \(T_{\mathrm{nested}}\,[\mathrm{s}]\) &
    Runtime ratio \\
    \hline
    \(20^2\)  &   3362 &    50.96 &    60.62 &  1.19 \\
    \(64^2\)  &  33282 &   301.78 &   695.08 &  2.30 \\
    \(128^2\) & 132098 &  1219.55 &  7260.85 &  5.95 \\
    \(256^2\) & 526338 &  5529.84 & 98164.47 & 17.75 \\
    \hline
  \end{tabular}
\end{table}

The computational advantage increases rapidly with problem size. For the
smallest benchmark both formulations exhibit comparable runtimes. However, the
benefit becomes increasingly pronounced as the number of global degrees of
freedom grows. For the largest benchmark containing more than half a million
global displacement unknowns, the runtime is reduced from approximately
$9.8\times10^4$\,s to $5.5\times10^3$\,s, corresponding to a runtime
reduction factor of $17.8$.

This behaviour directly reflects the underlying algorithmic idea. Both
formulations perform identical local constitutive updates. The proposed
monolithically condensed formulation, however, assembles and factorizes a
substantially smaller global nonlinear system in every Newton iteration.
Consequently, the computational advantage increases with the problem size,
making the proposed approach particularly attractive for large-scale finite
element simulations.

\begin{figure}[tb]
\centering

\begin{subfigure}[t]{0.48\textwidth}
    \centering
    \includegraphics[width=\textwidth]{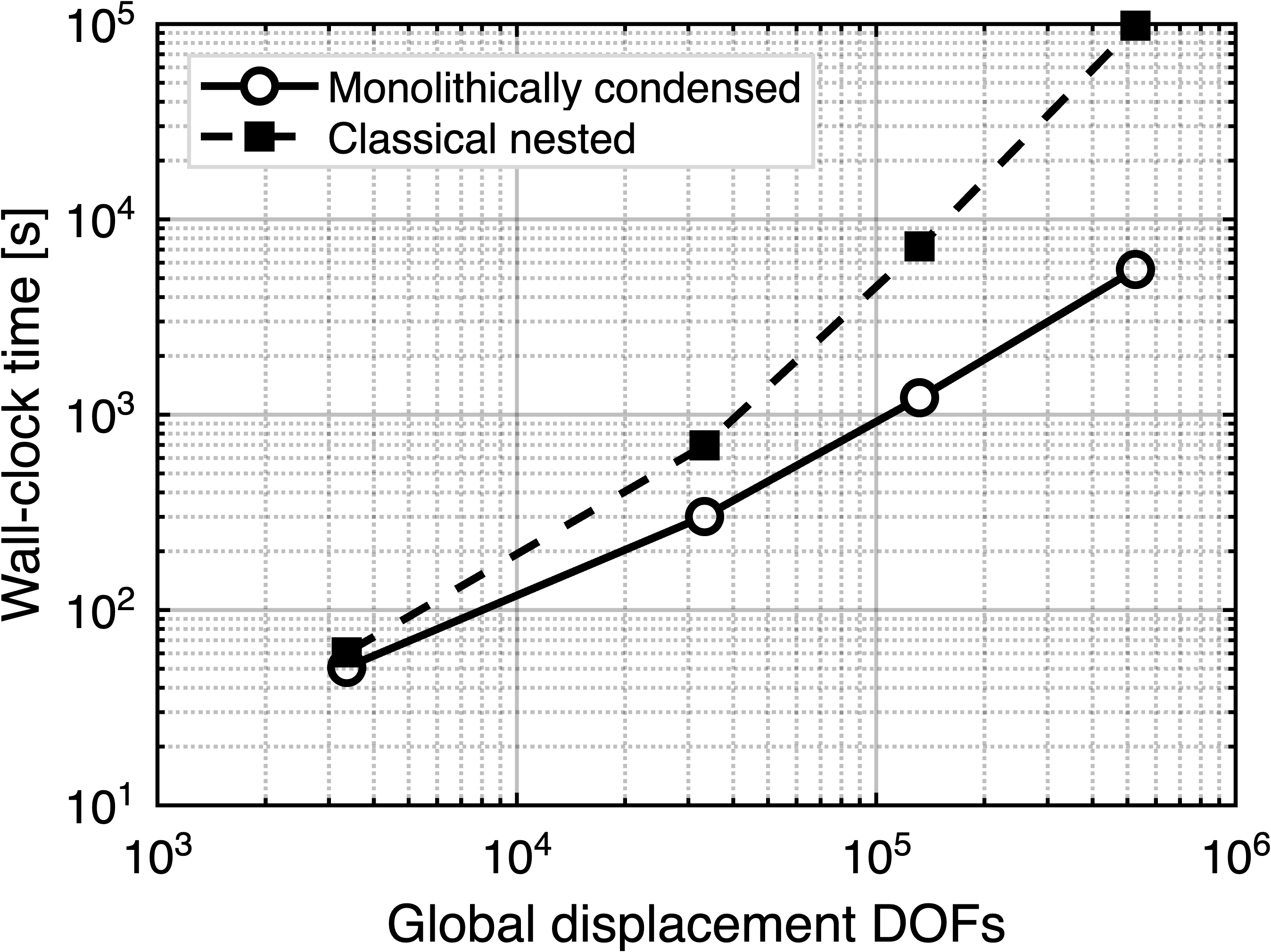}
    \caption{Wall-clock time for the classical nested and the proposed
    monolithically condensed formulation as a function of the number of
    global displacement degrees of freedom.}
    \label{fig:performance_runtime}
\end{subfigure}
\hfill
\begin{subfigure}[t]{0.48\textwidth}
    \centering
    \includegraphics[width=\textwidth]{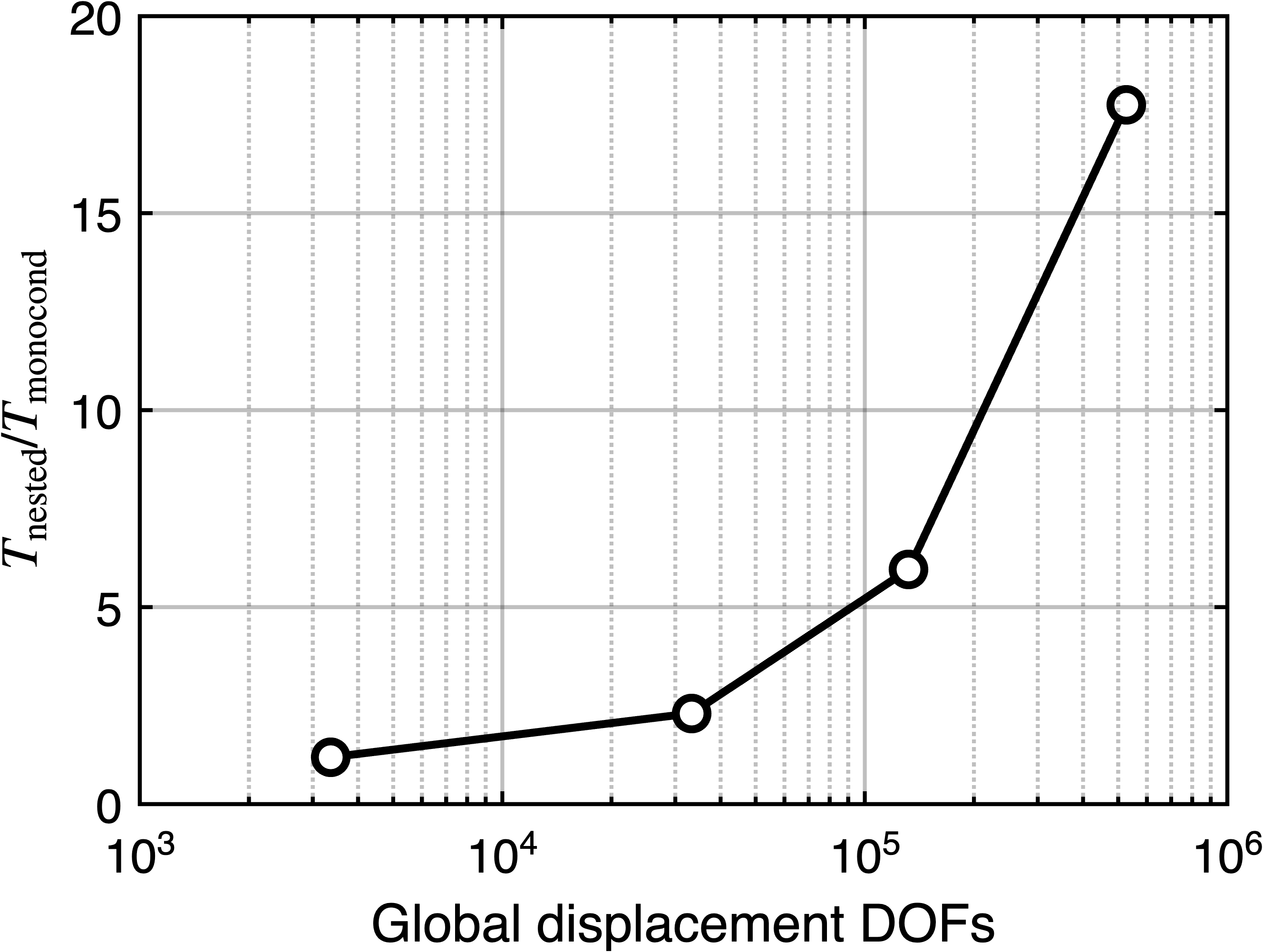}
    \caption{Runtime reduction factor
    $T_{\mathrm{nested}}/T_{\mathrm{monocond}}$
    versus the number of global displacement degrees of freedom.}
    \label{fig:performance_ratio}
\end{subfigure}

\caption{Computational performance of the proposed monolithically condensed
formulation for the Q2 Cook membrane benchmark. While both formulations solve
identical nonlinear problems, the computational advantage of the proposed
approach increases significantly with problem size. For the largest benchmark,
the runtime is reduced by a factor of approximately $17.8$.}
\label{fig:performance}
\end{figure}

\subsubsection{Influence of the nonlinear loading}

To investigate whether the observed performance gain depends on the degree of
nonlinearity, additional computations are performed on a fixed mesh using
three different load levels.

Only a moderate increase in runtime is observed with increasing load level.
This indicates that the additional cost associated with the nonlinear
constitutive response is comparatively small. Instead, the dominant
computational cost originates from assembling and solving the global linearized
systems. Consequently, the performance gain of the proposed formulation is
primarily governed by the reduction of the global problem size rather than by
changes in the nonlinear iteration counts.

Overall, the performance study demonstrates that the proposed formulation not
only preserves the solution accuracy established in
Section~\ref{sec:example2D}, but also provides substantial computational
savings that become increasingly significant for large-scale simulations.

\begin{table}[tb]
  \centering
  \caption{Influence of the load multiplier on the wall-clock time for the
  \(20\times20\) Q1 discretization. Each entry is the median of three
  independent runs.}
  \label{tab:sec52_load}
  \begin{tabular}{c r r r}
    \hline
    \(p\) &
    \(T_{\mathrm{monocond}}\,[\mathrm{s}]\) &
    \(T_{\mathrm{nested}}\,[\mathrm{s}]\) &
    \(T_{\mathrm{nested}}/T_{\mathrm{monocond}}\) \\
    \hline
    1 & 14.89 & 14.92 & 1.00 \\
    2 & 14.88 & 15.26 & 1.03 \\
    4 & 15.98 & 16.35 & 1.02 \\
    \hline
  \end{tabular}
\end{table}

\subsection{Discretization studies}\label{sec:conv_study}

The following studies separate the influence of the temporal discretization,
the approximation order of the displacement field, and the discontinuous
approximation chosen for the internal variable. All computations use the 2D
Cook's membrane introduced in Section~\ref{sec:example2D} with the reduced
load multiplier \(p=0.5\). Unless stated otherwise, the proposed
monolithically condensed formulation is employed.

\subsubsection{Temporal discretization}

The temporal sensitivity is investigated on a fixed
\(32\times32\) Q2 mesh using
\(n_t=10,20,40,80,160,320,\) and \(640\)
equal time steps on the interval
\([0,10]\).
The nonlinear tolerance is chosen as
\(10^{-8}\),
and all computations converge successfully.

The computation with
\(n_t=640\)
is used as the reference solution,
\[
  \mathbf u_{\mathrm{tip,ref}}
  =
  (-0.188973,\;0.148057),
  \qquad
  \|\mathbf u_{\mathrm{tip,ref}}\|
  =
  0.240066,
\]
and the relative temporal error is defined as
\[
e_t
=
\frac{
\|
\mathbf u_{\mathrm{tip}}^{\,n_t}
-
\mathbf u_{\mathrm{tip,ref}}
\|
}
{
\|
\mathbf u_{\mathrm{tip,ref}}
\|
}.
\]

The complete refinement sequence is summarized in
Table~\ref{tab:sec53_temporal}.

\begin{table}[tb]
\centering
\caption{Temporal refinement on a fixed \(32\times32\) Q2 mesh for
\(p=0.5\). The computation with \(n_t=640\) is used as the reference.}
\label{tab:sec53_temporal}
\begin{tabular}{r c c c c c c}
\hline
\(n_t\) &
\(\Delta t\) &
\(u_{\mathrm{tip},x}\) &
\(u_{\mathrm{tip},y}\) &
\(\|u_{\mathrm{tip}}\|\) &
\(e_t\) &
avg. Newton\\
\hline
10  & 1.00000  & -0.189068 & 0.148238 & 0.240252 & 8.50e-04 & 4.30\\
20  & 0.50000  & -0.189046 & 0.148169 & 0.240193 & 5.58e-04 & 3.95\\
40  & 0.25000  & -0.189040 & 0.148106 & 0.240149 & 3.48e-04 & 3.05\\
80  & 0.12500  & -0.188920 & 0.148034 & 0.240010 & 2.40e-04 & 3.00\\
160 & 0.06250  & -0.188888 & 0.148010 & 0.239970 & 4.04e-04 & 2.94\\
320 & 0.03125  & -0.188942 & 0.148041 & 0.240031 & 1.46e-04 & 2.00\\
640 & 0.015625 & -0.188973 & 0.148057 & 0.240066 & --        & 2.00\\
\hline
\end{tabular}
\end{table}

The temporal refinement exhibits a clear overall reduction of the
difference with respect to the reference solution. From
\(n_t=10\) to \(n_t=80\),
the relative tip error decreases almost monotonically by approximately
one order of magnitude. A small non-monotone deviation is observed for
the computation with \(n_t=160\), whose error is slightly larger than
that of the neighbouring refinement levels. Considering the strongly
nonlinear finite-deformation response of the Cook benchmark and the
small magnitude of the remaining differences (all below
\(10^{-3}\)), such a local deviation is not unexpected and does not
affect the overall convergence trend.

Most importantly, the temporal sensitivity is already substantially
smaller than the spatial discretization errors investigated below.
Consequently, the choice
\(n_t=80\),
used throughout the subsequent spatial and internal-variable studies,
provides sufficient temporal resolution for the present benchmark.

\subsubsection{Spatial convergence of Q1 and Q2 approximations}

For the spatial study, the time interval is divided into \(80\) equal steps.
The internal variable is represented by the quadrature-point-localized
\(\delta\) approximation. The Q1 sequence uses
\[
  8^2,\;16^2,\;32^2,\;64^2,\;128^2,\;256^2
\]
elements, whereas the Q2 sequence uses
\[
  4^2,\;8^2,\;16^2,\;32^2,\;64^2,\;128^2,\;256^2
\]
elements. All thirteen computations converged without damping difficulties,
with average Newton iteration counts between \(2.95\) and \(3.00\).

As a provisional common reference, we use the finest Q2 result,
\[
  \vec u_{\mathrm{tip,ref}}
  =
  (-0.190146,\;0.148421),
  \qquad
  \|\vec u_{\mathrm{tip,ref}}\|
  =
  0.241214.
\]
The relative tip error is defined as
\[
  e_{\mathrm{tip}}
  =
  \frac{\|\vec u_{\mathrm{tip}}
  -\vec u_{\mathrm{tip,ref}}\|}
  {\|\vec u_{\mathrm{tip,ref}}\|}.
\]
Table~\ref{tab:sec53_spatial} summarizes the complete refinement sequence.

\begin{table}[tb]
  \centering
  \caption{Spatial refinement of the 2D Cook's membrane for \(p=0.5\) and
  \(n_t=80\). The finest Q2 result serves as the provisional reference.
  \(N_u\) denotes the number of globally coupled displacement unknowns.}
  \label{tab:sec53_spatial}
  \begin{tabular}{c r r c c c}
    \hline
    Order & Elements/dir. & \(N_u\) & \(\|u_{\mathrm{tip}}\|\)
          & \(e_{\mathrm{tip}}\) & rate \\
    \hline
1 & 8 & 162 & 0.183954 & 2.499e-01 & -- \\
1 & 16 & 578 & 0.218236 & 1.015e-01 & 1.30 \\
1 & 32 & 2178 & 0.232341 & 3.958e-02 & 1.36 \\
1 & 64 & 8450 & 0.237611 & 1.616e-02 & 1.29 \\
1 & 128 & 33282 & 0.239647 & 7.013e-03 & 1.20 \\
1 & 256 & 132098 & 0.240510 & 3.120e-03 & 1.17 \\
2 & 4 & 162 & 0.228395 & 5.593e-02 & -- \\
2 & 8 & 578 & 0.235530 & 2.534e-02 & 1.14 \\
2 & 16 & 2178 & 0.238677 & 1.132e-02 & 1.16 \\
2 & 32 & 8450 & 0.240010 & 5.331e-03 & 1.09 \\
2 & 64 & 33282 & 0.240656 & 2.444e-03 & 1.12 \\
2 & 128 & 132098 & 0.241007 & 8.976e-04 & 1.45 \\
2 & 256 & 526338 & 0.241214 & -- & -- \\
    \hline
  \end{tabular}
\end{table}

The error decreases monotonically for both approximation orders. Over most of
the Q1 sequence, the observed rate lies between approximately \(1.2\) and
\(1.36\), while the Q2 sequence exhibits rates close to \(1.1\) over the first
five refinement levels. The reduced rates are consistent with the limited
regularity of the Cook benchmark caused by the abrupt transition from the
clamped to the free boundary. The purpose of the study is therefore not to
recover the formal interpolation order of a smooth problem, but to demonstrate
systematic convergence and the increased efficiency of the higher-order
approximation.

The comparison at equal numbers of global displacement unknowns is especially
instructive. Q1 meshes with \(8,16,32,64,\) and \(128\) elements per direction
have essentially the same \(N_u\) as Q2 meshes with \(4,8,16,32,\) and \(64\)
elements per direction, respectively. At each of these pairs, Q2 yields a
substantially smaller tip error. For example, at \(N_u=33282\), the relative
error decreases from \(7.01\times10^{-3}\) for Q1 to
\(2.44\times10^{-3}\) for Q2.

\subsubsection{Approximation space of the internal variable}

The final study compares three discontinuous approximations of the
strain-like internal variable on a common Q2 displacement discretization,
\[
  \mathcal C_i^h \in
  \left\{
    \operatorname{span}\{\delta_g\},
    P_0^{\mathrm{disc}},
    Q_1^{\mathrm{disc}}
  \right\}.
\]
Here, \(\delta_g\) denotes the quadrature-point-localized representation,
\(P_0^{\mathrm{disc}}\) an element-wise constant field, and
\(Q_1^{\mathrm{disc}}\) an element-wise bilinear field. Since this study
concerns the approximation properties of the internal field rather than the
performance of the nonlinear solution strategies, all cases are computed
with the fully monolithic formulation, which supports the three spaces
without changing the underlying weak form. The displacement field is
quadratic and the time interval is discretized by \(n_t=80\) equal steps.

The finest quadrature-point-localized result on the \(64\times64\) mesh is
used as a common reference,
\[
  \vec u_{\mathrm{tip,ref}}
  =
  (-0.189590,\;0.148226),
  \qquad
  \|\vec u_{\mathrm{tip,ref}}\|=0.240656.
\]
Table~\ref{tab:sec53_internal} reports the number \(N_C\) of discrete
internal-variable unknowns, the final tip-displacement norm, the relative
tip error with respect to this reference, and the measured wall-clock time.

\begin{table}[tb]
  \centering
  \caption{Influence of the discontinuous approximation space of the
  internal variable for a Q2 displacement field and \(n_t=80\). The finest
  quadrature-point-localized result is used as the common reference.}
  \label{tab:sec53_internal}
  \begin{tabular}{c r r c c r}
    \hline
    Internal space & Elements/dir. & \(N_C\)
    & \(\|u_{\mathrm{tip}}\|\) & rel. error & time [s] \\
    \hline
$\delta_g$ & 4 & 576 & 0.228395 & 5.361e-02 & 20.3 \\
$\delta_g$ & 8 & 2304 & 0.235530 & 2.295e-02 & 78.4 \\
$\delta_g$ & 16 & 9216 & 0.238677 & 8.903e-03 & 240.8 \\
$\delta_g$ & 32 & 36864 & 0.240010 & 2.895e-03 & 411.1 \\
$\delta_g$ & 64 & 147456 & 0.240656 & -- & 1602.4 \\
$P_0^{\mathrm{disc}}$ & 4 & 64 & 0.212010 & 1.247e-01 & 15.7 \\
$P_0^{\mathrm{disc}}$ & 8 & 256 & 0.228379 & 5.408e-02 & 47.0 \\
$P_0^{\mathrm{disc}}$ & 16 & 1024 & 0.235466 & 2.313e-02 & 185.5 \\
$P_0^{\mathrm{disc}}$ & 32 & 4096 & 0.238586 & 9.311e-03 & 192.7 \\
$P_0^{\mathrm{disc}}$ & 64 & 16384 & 0.240000 & 2.992e-03 & 556.2 \\
$Q_1^{\mathrm{disc}}$ & 4 & 256 & 0.225489 & 6.594e-02 & 14.4 \\
$Q_1^{\mathrm{disc}}$ & 8 & 1024 & 0.234211 & 2.875e-02 & 52.8 \\
$Q_1^{\mathrm{disc}}$ & 16 & 4096 & 0.238113 & 1.143e-02 & 209.3 \\
$Q_1^{\mathrm{disc}}$ & 32 & 16384 & 0.239754 & 4.047e-03 & 271.6 \\
$Q_1^{\mathrm{disc}}$ & 64 & 65536 & 0.240530 & 5.648e-04 & 929.4 \\
    \hline
  \end{tabular}
\end{table}

All three approximation families approach the same response under mesh
refinement. On the finest mesh, the relative difference from the
quadrature-point reference is \(5.65\times10^{-4}\) for
\(Q_1^{\mathrm{disc}}\) and \(2.99\times10^{-3}\) for
\(P_0^{\mathrm{disc}}\). The corresponding reduction in the number of
internal unknowns is substantial. On the \(64\times64\) mesh,
\(P_0^{\mathrm{disc}}\) requires \(16384\) internal unknowns instead of
\(147456\) for the quadrature-point representation, i.e. a factor of nine
fewer, while the wall-clock time decreases from \(1602\) to \(556\) seconds.
The discontinuous Q1 approximation occupies an intermediate position with
\(65536\) internal unknowns, a relative difference below \(6\times10^{-4}\),
and a runtime of \(929\) seconds.

These results confirm the discrete interpretation developed in
Section~\ref{sec:spatial_FE}. The quadrature-point-localized representation is
one particular discontinuous approximation of the local evolution equation,
not an intrinsic requirement of the constitutive model or of the monolithic
solution strategy. Element-wise polynomial internal fields provide alternative projections
of the same local evolution law and may offer a favourable compromise
between accuracy and computational effort. The timing
values in Table~\ref{tab:sec53_internal} are included only to illustrate this
trade-off; they are not used for the solver-performance comparison of
Section~5.2 because the three approximation spaces contain different numbers
of internal unknowns.

\subsection{3D Cook's membrane}

To assess whether the observations from the previous two-dimensional
examples also carry over to a genuinely three-dimensional setting, we
consider the classical 3D Cook's membrane benchmark. The reference
configuration is defined by the corner points
\[
P_1=(0,0,0),\;
P_2=(480,440,0),\;
P_3=(480,600,0),\;
P_4=(0,440,0),
\]
\[
P_5=(0,0,160),\;
P_6=(480,440,160),\;
P_7=(480,600,160),\;
P_8=(0,440,160),
\]
where all coordinates are given in mm.

The left face
\[
\Gamma_u
=
\{
X\in\mathbb{R}^3
\mid
X_1=0,\;
0<X_2<440,\;
0<X_3<160
\}
\]
is fixed by homogeneous Dirichlet boundary conditions. On the opposite
face
\[
\Gamma_t
=
\{
X\in\mathbb{R}^3
\mid
X_1=480,\;
440<X_2<600,\;
0<X_3<160
\},
\]
the time-dependent traction
\[
\mathbf T(X,t)
=
\Lambda(t)\,
p\,
(-750,\;1000,\;500)^T
\]
is prescribed, where $p$ denotes the dimensionless load multiplier.
Body forces are neglected.

The same viscoelastic material model and constitutive parameters as in
the previous examples are employed. The displacement field is
interpolated by trilinear eight-node hexahedral elements on a
$20\times20\times5$ mesh. The internal variables are stored at the
quadrature points, resulting in a block-diagonal structure of the local
internal tangent. The time interval $[0,10]$ is discretized by
thirteen equal time steps.

Since the computational efficiency has already been assessed in Section \ref{sec:performance}, the objective of the present example is solely to investigate nonlinear robustness. To this end, the load
increment is successively increased by reducing the number of time
steps. For every load multiplier, the minimum number of time steps
required for successful convergence is determined.

\begin{table}[tb]
\centering
\caption{Minimum number of time steps required for convergence in the
3D Cook's membrane benchmark.}
\label{tab:sec54_threshold}
\begin{tabular}{cccc}
\hline
Load multiplier &
Monolithically condensed &
Classical nested &
Ratio
\\
$p$
&
$n_{\min}^{\mathrm{mono}}$
&
$n_{\min}^{\mathrm{nested}}$
&
$n_{\min}^{\mathrm{nested}}/
 n_{\min}^{\mathrm{mono}}$
\\
\hline
0.5 & 3 & 3 & 1.00\\
1.0 & 5 & 5 & 1.00\\
1.5 & 8 & 13 & 1.63\\
2.0 & 9 & 13 & 1.44\\
\hline
\end{tabular}
\end{table}

For the two smaller load levels, both nonlinear solution strategies
exhibit essentially identical robustness. In particular, both methods
converge with three and five time steps for $p=0.5$ and $p=1.0$,
respectively. Hence, under moderate loading, the proposed monolithic
condensation reproduces the robustness of the classical nested
Gauss-point procedure.

A clear difference appears for the larger load levels. For
$p=1.5$, the proposed monolithically condensed formulation converges
already with eight time steps, whereas the classical nested strategy
requires thirteen. Likewise, for $p=2.0$, the proposed method converges
with nine time steps, while the nested formulation again requires
thirteen. Consequently, the admissible load increment can be increased
by approximately $63\,\%$ for $p=1.5$ and by approximately $44\,\%$ for
$p=2.0$.

These observations complement the previous sections. Section~5.1
demonstrated that the proposed formulation is algebraically equivalent
to the fully monolithic system, while Section~5.2 showed that the
condensation significantly reduces the computational cost. The present
three-dimensional example indicates that retaining the internal
residuals within the global Newton process may additionally improve the
nonlinear robustness for challenging load increments without changing
the underlying constitutive model or its spatial discretization.

\subsection{Nonlinear multi-branch viscoelasticity}\label{sec:powerlaw_multibranch}

The preceding examples deliberately use a single linear viscous branch in order to keep
both the constitutive setting and the comparison between the nonlinear solution strategies
transparent. To demonstrate that the proposed monolithic condensation is not tied to this
particular viscosity law, we additionally consider a nonlinear multi-branch extension of
the same finite-viscoelastic structure. The purpose of this example is algorithmic rather
than constitutive calibration: it shows that the same block structure and the same
Schur/null-space condensation remain applicable when several internal tensors and a
stress-dependent power-law-type evolution equation are used.

The free energy is written as
\begin{equation}
  \Psi\bigl(C,\{\mathcal C_a\}_{a=1}^{N_b}\bigr)
  =
  \Psi_{\infty}(C)
  +
  \sum_{a=1}^{N_b}\Psi_a(C_{e,a}),
  \qquad
  C_{e,a}=C\mathcal C_a^{-1},
  \label{eq:powerlaw_free_energy}
\end{equation}
where \(\mathcal C_a\) denotes the strain-like internal variable of branch \(a\). Each
non-equilibrium branch uses the same functional form as in Section~\ref{sec:example2D},
but with branch-weighted moduli
\(\mu_a=g_a\mu_{\mathrm{visc}}\) and
\(\lambda_a=g_a\lambda_{\mathrm{visc}}\). The branch driving force is
\begin{equation}
  M_a
  =
  2\frac{\partial\Psi_a}{\partial C_{e,a}}\mathcal C_a^{-1}C .
  \label{eq:powerlaw_driving_force}
\end{equation}
Instead of the linear mobility used in Section~2.4, we prescribe the evolution equation
in the form
\begin{equation}
  \dot{\mathcal C}_a
  =
  2\mathcal C_a A_a(M_a),
  \label{eq:powerlaw_evolution}
\end{equation}
with
\begin{equation}
  A_a(M_a)
  =
  \alpha_{\mathrm{vol},a}\,\operatorname{vol}(M_a^T)
  +
  \alpha_{\mathrm{dev},a}\,
  \chi_a\bigl(\|\operatorname{dev}(M_a^T)\|\bigr)
  \operatorname{dev}(M_a^T),
  \label{eq:powerlaw_mobility}
\end{equation}
and
\begin{equation}
  \chi_a(r)
  =
  \left(
    1 + \frac{r^2}{M_{0,a}^2}
  \right)^{(m_a-1)/2} .
  \label{eq:powerlaw_chi}
\end{equation}
Here, \(\alpha_{\mathrm{vol},a}>0\), \(\alpha_{\mathrm{dev},a}>0\), and \(M_{0,a}>0\)
are branch parameters. In the computations below we use
\(\alpha_{\mathrm{dev},a}=1/(2V_{\mathrm{dev},a})\) and
\(\alpha_{\mathrm{vol},a}=1/(dV_{\mathrm{vol},a})\), with \(d=3\) for the plane-strain
constitutive embedding. For \(m_a=1\), \(\chi_a=1\) and the linear branch law is
recovered. For \(m_a>1\), the deviatoric mobility increases with the magnitude of the
thermodynamic driving force and produces a power-law-type internal evolution. Moreover,
for each branch,
\begin{equation}
  D_a
  =
  M_a^T:A_a(M_a)
  =
  \alpha_{\mathrm{vol},a}\|\operatorname{vol}(M_a^T)\|^2
  +
  \alpha_{\mathrm{dev},a}\chi_a\|\operatorname{dev}(M_a^T)\|^2
  \ge 0,
\end{equation}
so that the Clausius--Duhem inequality is preserved.

For the numerical test we use \(N_b=3\) branches with the parameters summarized in
Table~\ref{tab:powerlaw_branch_params}. The branch weights satisfy \(\sum_a g_a=1\).
The viscosities span two orders of magnitude and therefore represent a spectrum of
nominal relaxation times. The exponent \(m_a=3\) gives a cubic stress-dependent
contribution to the deviatoric mobility. We also run a reference case with
\(m_a=1\), which recovers the corresponding linear three-branch model.

\begin{table}[h!]
  \centering
  \caption{Branch parameters for the nonlinear multi-branch material.}
  \label{tab:powerlaw_branch_params}
  \begin{tabular}{|c|c|c|c|c|c|}
    \hline
    Branch $a$ & $g_a$ & $V_{\mathrm{dev},a}$ & $V_{\mathrm{vol},a}$ & $M_{0,a}$ & $m_a$ \\
    \hline
    1 & 0.5 & $1.0\times 10^3$ & $5.0\times 10^3$ & $1.5\times 10^4$ & 3 \\
    2 & 0.3 & $1.0\times 10^4$ & $5.0\times 10^4$ & $1.5\times 10^4$ & 3 \\
    3 & 0.2 & $1.0\times 10^5$ & $5.0\times 10^5$ & $1.5\times 10^4$ & 3 \\
    \hline
  \end{tabular}
\end{table}

The benchmark geometry, boundary conditions, loading direction, displacement
interpolation, and internal-variable storage are identical to the 2D Cook's membrane in
Section~\ref{sec:example2D}. We use a \(20\times20\) Q1 mesh, four integration points per
element, and \(n=10\) time steps on \([0,10]\,\mathrm{s}\). The internal variable is stored
at the quadrature points. Since three branches are used, the number of stored internal
unknowns is tripled compared with the single-branch example. Both the monolithic and the
classical nested condensation strategies use the same spatial and temporal discretization;
the only difference is again whether the internal equations are retained in the global
Newton residual and condensed at the linearized level, or solved by nested local Newton
iterations at the quadrature points.

As a first consistency check, the linear three-branch reference with \(m_a=1\) was run for
\(p=1\). The final tip displacement norm was
\(\|u_{\mathrm{tip}}\|=3.5330539\times10^{-1}\) for both strategies, with a difference
below \(7\times10^{-12}\). The average number of outer Newton iterations was 4.7 for the
monolithic strategy and 4.4 for the classical nested condensation; the corresponding
wall-clock times were 23.83s and 42.33s.

Table~\ref{tab:powerlaw2d_newton_times} reports the results for the nonlinear
three-branch model with \(m_a=3\). For all load multipliers, both methods converge and
yield the same mechanical end state up to round-off accuracy. The differences in the
final tip displacement between monolithic and nested condensation are
\(6.43\times10^{-12}\), \(4.16\times10^{-12}\), and \(9.10\times10^{-12}\) for
\(p=1,2,4\), respectively. The outer Newton iteration counts remain comparable. The
computational cost, however, is significantly smaller for the monolithic strategy because
it avoids the nested local Newton solves. The resulting speed-up factors are approximately
1.89, 1.97, and 1.69 for \(p=1,2,4\), respectively.

\begin{table}[h!]
  \centering
  \caption{Nonlinear three-branch material with power-law mobility: average outer Newton iterations and wall-clock times for the 2D Cook's membrane.}
  \label{tab:powerlaw2d_newton_times}
  \begin{tabular}{|c|cc|cc|cc|}
    \hline
    & \multicolumn{2}{c|}{$p=1$} & \multicolumn{2}{c|}{$p=2$} & \multicolumn{2}{c|}{$p=4$} \\
    \cline{2-7}
    Solution strategy
    & $n_{\mathrm{it}}$ & time [s]
    & $n_{\mathrm{it}}$ & time [s]
    & $n_{\mathrm{it}}$ & time [s] \\
    \hline
    Monolithic & 4.7 & 23.38 & 4.9 & 23.59 & 6.1 & 32.78 \\
    \hline
    Classical (nested condensation) & 4.4 & 44.21 & 4.6 & 46.55 & 5.3 & 55.43 \\
    \hline
  \end{tabular}
\end{table}

Finally, we repeat the ramp-and-hold test with the nonlinear three-branch model for
\(p=1\). The traction is ramped during \(0\le t\le10\,\mathrm{s}\) and then kept constant
until \(t=15\,\mathrm{s}\). The results are shown in Table~\ref{tab:powerlaw2d_creep}.
Again, both strategies produce the same final response; the difference in the final tip
displacement is \(1.07\times10^{-11}\). The monolithic computation requires about half
the wall-clock time of the classical nested condensation.

\begin{table}[h!]
  \centering
  \caption{Ramp-and-hold response for the nonlinear three-branch material with $p=1$.}
  \label{tab:powerlaw2d_creep}
  \begin{tabular}{|c|c|c|c|c|}
    \hline
    Solution strategy & Avg. Newton & Time [s] & $u_{\mathrm{tip},x}$ & $u_{\mathrm{tip},y}$ \\
    \hline
    Monolithic condensed & 4.33 & 31.94 & $-3.1129\times10^{-1}$ & $2.0156\times10^{-1}$ \\
    \hline
    Classical (nested condensation) & 4.00 & 60.94 & $-3.1129\times10^{-1}$ & $2.0156\times10^{-1}$ \\
    \hline
  \end{tabular}
\end{table}

This example confirms the algorithmic generality of the proposed condensation: the
monolithic reduction is not restricted to a single linear Maxwell-type branch. It applies
unchanged to several internal tensors and to a nonlinear, stress-dependent evolution law.
Only the internal residual block and its consistent tangent are enlarged; the global block
structure and the Schur/null-space condensation remain the same.

\section{Conclusions}

We have presented a fully discrete monolithic solution framework for finite
viscoelasticity with a strain-like internal variable. Rather than treating
the constitutive update as an intrinsically separate material-point problem,
the deformation and the internal state are regarded as unknowns of one
coupled nonlinear algebraic system. Its Newton linearization yields a
naturally non-symmetric block tangent. The internal-variable increment can
then be eliminated consistently by a Schur complement and reconstructed
after the reduced displacement solve. The same reduction admits a geometric
interpretation in terms of a null-space basis of the tangent space of the
time-discrete internal constraint manifold. The classical nested
Gauss-point update is thereby identified as one particular nonlinear
solution strategy for the same fully discrete problem, distinguished by
enforcing the internal residual locally before each global Newton step.

The numerical studies support this interpretation from several complementary
perspectives. The fully monolithic, monolithically condensed, and classical
nested implementations converge to the same discrete solution, while the
proposed condensation reduces the globally coupled system to the displacement
unknowns without degrading the outer Newton convergence. In the Q2 scaling
study, the computational advantage grows markedly with problem size: for the
largest two-dimensional benchmark with more than half a million global
displacement unknowns, the measured wall-clock time is reduced by a factor of
approximately \(17.8\). The temporal and spatial refinement studies confirm
that the solver comparisons are not governed by an unresolved discretization
error. Moreover, the internal-variable study shows that the
quadrature-point-localized representation is only one member of a broader
family of discontinuous approximations. Element-wise constant and bilinear
internal fields approach the same response while substantially reducing the
number of internal unknowns; for the finest mesh considered, the
element-wise constant approximation uses a factor of nine fewer internal
unknowns than the quadrature-point representation.

The three-dimensional Cook benchmark additionally reveals a robustness
advantage for challenging load increments. For moderate load levels, both
solution strategies exhibit the same convergence threshold. At the larger
load multipliers, however, the monolithically condensed formulation remains
convergent with eight and nine time steps, whereas the classical nested
scheme requires thirteen. Retaining the non-vanishing internal residual in
the condensed global Newton correction can therefore permit substantially
larger load increments. Finally, the nonlinear three-branch example confirms
that the construction is not restricted to a single linear Maxwell-type
branch: the block structure and condensation remain unchanged for several
internal tensors and a stress-dependent power-law mobility, while the
repeated nonlinear material-point solves of the classical procedure are
avoided.

The contribution is therefore primarily algorithmic. It does not propose a
new constitutive law, but recasts the solution architecture of
internal-variable models at the fully discrete level. The construction
requires an internal residual, its consistent tangent, and local
invertibility of the internal block---the same regularity needed by the
classical constitutive update. This makes the approach directly relevant to
broader classes of thermodynamically consistent internal-variable models,
including generalized standard materials, viscoplastic regularizations, and
computational multiscale formulations in which nested local nonlinear solves
are particularly expensive. Future work will address scalable iterative
solvers and preconditioners for the resulting non-symmetric systems, discrete
energy--dissipation verification, and applications to more demanding
three-dimensional and multiscale problems.

\section*{Acknowledgments}
Support for this research was provided by the Deutsche Forschungsgemeinschaft
(DFG), Germany under grant HE5943/24-1 and HE5943/26-1. This support is
gratefully acknowledged.


\bibliographystyle{plain}
\bibliography{literature}

\end{document}